\pdfoutput=1
\documentclass[11pt, letterpaper, logo, onecolumn, copyright]{googledeepmind}

\usepackage[authoryear, sort&compress, round]{natbib}
\usepackage[most, breakable, skins]{tcolorbox}
\tcbuselibrary{skins}
\usepackage{tabularx}
\usepackage{subcaption}
\usepackage{multirow}
\usepackage{multicol}
\usepackage{array}
\usepackage{amssymb,graphicx}
\usepackage{booktabs}
\usepackage{longtable}
\usepackage{newfloat}
\usepackage{float}
\usepackage{enumitem}
\usepackage{arydshln}
\usepackage[dvipsnames]{xcolor}
\definecolor{bestrow}{RGB}{235,245,255}
\definecolor{bestleft}{RGB}{235,245,255}
\definecolor{bestright}{RGB}{235,245,255}
\usepackage{hyperref}
\usepackage{CJKutf8}

\usepackage{xcolor}
\usepackage{listings}

\lstdefinelanguage{json}{
    basicstyle=\ttfamily\small,
    showstringspaces=false,
    breaklines=true,
    frame=none,
    columns=fullflexible,
    morestring=[b]",
    stringstyle=\color{red!70!black},
    literate=
     *{:}{{{\color{black}{:}}}}{1}
      {,}{{{\color{black}{,}}}}{1}
      {\{}{{{\color{black}{\{}}}}{1}
      {\}}{{{\color{black}{\}}}}}{1}
      {[}{{{\color{black}{[}}}}{1}
      {]}{{{\color{black}{]}}}}{1}
      {0}{{{\color{blue}0}}}{1}
      {1}{{{\color{blue}1}}}{1}
      {2}{{{\color{blue}2}}}{1}
      {3}{{{\color{blue}3}}}{1}
      {4}{{{\color{blue}4}}}{1}
      {5}{{{\color{blue}5}}}{1}
      {6}{{{\color{blue}6}}}{1}
      {7}{{{\color{blue}7}}}{1}
      {8}{{{\color{blue}8}}}{1}
      {9}{{{\color{blue}9}}}{1},
}

\lstdefinestyle{myminted}{
  basicstyle=\ttfamily\scriptsize,
  breaklines=true,
  columns=fullflexible,
  tabsize=2,
  frame=none,
  aboveskip=0pt,
  belowskip=0pt,
  xleftmargin=0pt,
  xrightmargin=0pt,
  lineskip=0pt,
  keepspaces=true,
}

\tcbuselibrary{listings}

\newcolumntype{L}[1]{>{\raggedright\let\newline\\\arraybackslash\hspace{0pt}}m{#1}}
\newcolumntype{C}[1]{>{\centering}m{#1}}
\newcolumntype{R}[1]{>{\raggedleft\let\newline\\\arraybackslash\hspace{0pt}}m{#1}}

\definecolor{ao}{rgb}{0.0, 0.0, 1.0}

\let\cite\citep

\DeclareFloatingEnvironment[
  fileext=loc,
  listname={List of Code Listings},
  name=Listing,
  placement=H
]{codelisting}
\makeatletter
\renewenvironment{codelisting}[1][]{%
  \par\addvspace{\medskipamount}%
  \def\@captype{codelisting}%
  \captionsetup{hypcap=false}%
}{%
  \par\addvspace{\medskipamount}%
}
\makeatother

\tcbset{
  colback=black!2,
  colframe=black!15,
  arc=1mm,
  boxrule=0.3mm,
  left=2mm,
  right=2mm,
  top=1mm,
  bottom=1mm
}

\title{Prompt Generation Technical Report}
\reportnumber{}
\renewcommand{\today}{}

\author[1]{\large Taobao Search Team}

\newcommand{\eg}{\textit{e.g.}}
\newcommand{\ie}{\textit{i.e.}}

\begin{abstract}
Generative retrieval has become an increasingly adopted paradigm for industrial search, recommendation, and advertising systems, delivering significant online gains. Most existing work combines user behavior sequences with large language models (LLMs) to model user preferences. In practice, feature engineering remains critical to model effectiveness, yet its complexity slows offline iteration and makes online deployment heavy and hard to reuse, all under tight online latency budgets. The root cause is a tight coupling between feature-processing logic and model architecture, where every feature change touches the training and serving code and resists reuse across scenarios.

To break this coupling, as shown in Figure~\ref{fig:pg_abstract}, we present \textit{Prompt Generation (PG)}, a high-level tokenizer and configuration-driven framework that decouples feature-processing logic from model architecture through two declarative JSON files, which serve as the single source of truth for both offline training and online serving, ensuring feature consistency across the two stages. Organizing features under four types with three composable processing components to assemble and compress heterogeneous features, PG delivers acceleration at three levels:\\
\hspace*{1em} \noindent\ (1)~\textit{fast training iteration}: feature experiments require only configuration changes, with built-in \hspace*{3em} token compression for ultra-long sequences;\\
\hspace*{0.9em} \noindent\ (2)~\textit{fast deployment}: a new scenario only needs to conform to the PG schema and plug into a \hspace*{3em} universal pipeline, with no scenario-specific engineering;\\
\hspace*{1.1em} \noindent\  (3)~\textit{fast online inference}:  engine applies unified optimizations over the standardized configuration, \hspace*{3.2em}reducing PG's overhead to a negligible level.\\
We validate PG on Taobao Search, Taobao Recommendation, and three open-source benchmarks, where configuration adaptation alone improves offline retrieval quality across all five scenarios. In production, PG has been deployed on Taobao Search with statistically significant online A/B uplifts of $+0.47\%$ in transaction count and $+0.51\%$ in GMV, and has been applied across multiple Taobao search and recommendation teams as the iteration framework for generative retrieval.
\end{abstract}

\begin{document}

\maketitle

\begin{figure}[H]
    \centering
    \includegraphics[width=\textwidth]{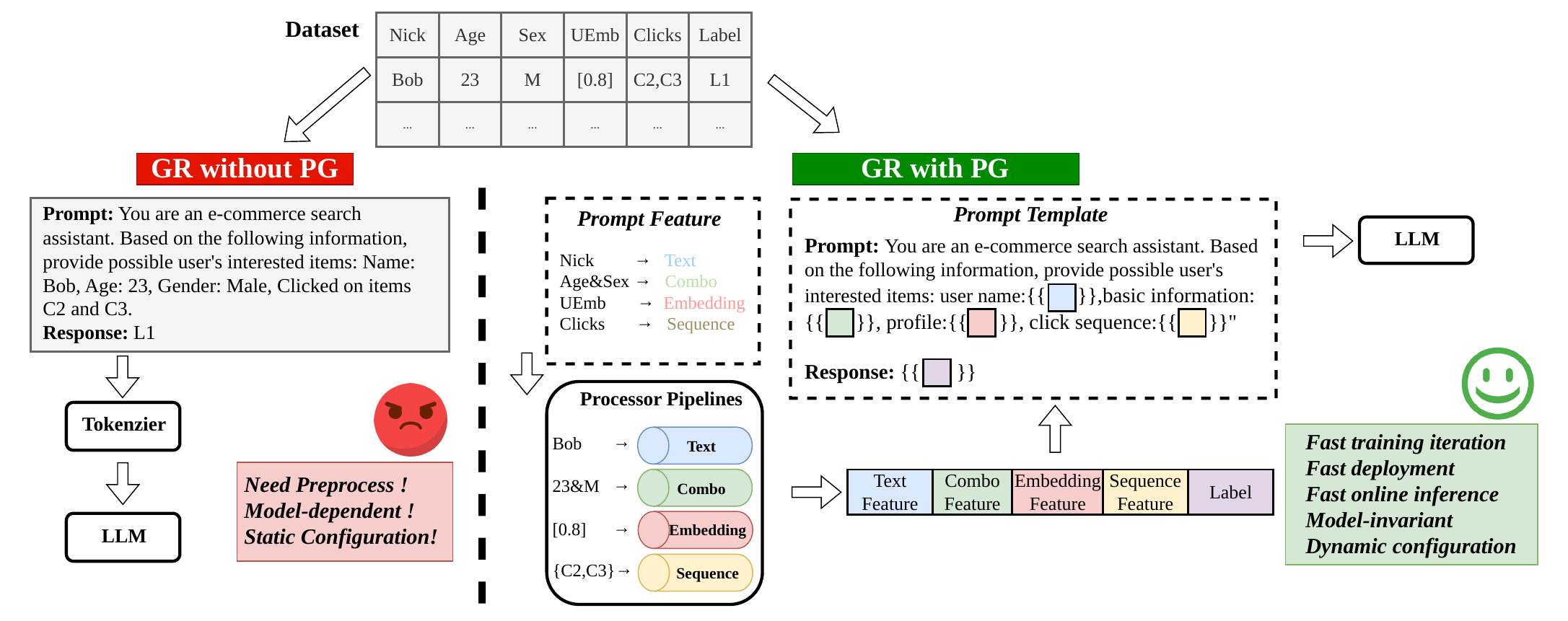}
    \caption{Generative Retrieval (GR) without vs. with Prompt Generation (PG). PG decouples feature processing from model architecture, turning a static pipeline into a dynamic, model-agnostic one.}
    \label{fig:pg_abstract}
\end{figure}

\newpage
\tableofcontents
\newpage

\section{Introduction}
\label{sec:introduction}

Large language models (LLMs) have demonstrated remarkable capabilities in understanding and generating natural language, leading to their increasing adoption in information retrieval systems. In particular, LLMs have shown great promise in \emph{generative retrieval}, a paradigm shift from traditional retrieval methods that rank pre-existing candidates to directly generating target items (\eg, document identifiers, product IDs) based on input queries~\citep{tay2022transformer,bevilacqua2022autoregressive,li2025matching}. This generative approach has been successfully applied across diverse domains, including web search~\citep{metzler2021rethinking}, recommendation systems~\citep{rajput2023recommender}, and sponsored search advertising~\citep{kong2025generative}, demonstrating superior performance in capturing complex user intent and item relationships. The shift from traditional retrieval to generative retrieval represents a fundamental change in how systems process and utilize information. Rather than relying on explicit matching signals between queries and candidates, generative models learn to encode rich semantic relationships into their parameters and generate relevant results through sequence-to-sequence modeling.

However, deploying generative retrieval models in real-world production systems presents significant technical challenges, particularly in managing the heterogeneous features required for effective model input construction. Unlike traditional retrieval systems where feature engineering follows well-established patterns, LLM-based generative models must integrate diverse feature types, including raw text fields, pre-computed embeddings, combo features, and temporal behavior sequences, into carefully designed prompts that combine structured information with natural language context~\citep{fu2025forge,deng2025onerec}. Each feature type demands distinct processing pipelines, creating a complex interplay between feature representation and model input format. Manual implementation of such feature processing logic is not only error-prone but also fails to scale across different scenarios, as each new application domain or model architecture may necessitate substantial re-engineering efforts. Furthermore, production machine learning systems face the critical challenge of maintaining consistency between offline training and online serving environments, a phenomenon commonly known as training-serving skew, where discrepancies in feature computation can severely degrade model performance~\citep{bitton2023evaluating}. The absence of a systematic framework for configuring, validating, and deploying these heterogeneous features creates bottlenecks in both rapid experimentation during model development and reliable deployment in production environments.

To address these challenges, we propose a configuration-driven \textbf{Prompt Generation (PG)} framework that systematically manages heterogeneous features in LLM-based generative retrieval systems. Our approach introduces a unified protocol that supports four fundamental feature types commonly encountered in production systems, including raw text fields, pre-computed embeddings, combo features, and temporal behavior sequences. Built upon a JSON-based configuration schema, the framework allows for the declarative specification of feature processing logic. It incorporates three types of composable components: preprocessors for data transformation, projectors for representation mapping and compression, and mergers for feature combination. These components enable flexible feature manipulation, allowing developers to balance input complexity and computational cost while preserving essential information. By decoupling feature processing logic from the underlying model architecture, this design facilitates rapid iteration without requiring infrastructure modifications. 
We also provide comprehensive design guidelines for training frameworks and serving infrastructure.
To address the critical challenge of training-serving skew, we implement an event tracking mechanism. This system logs all source features from the online serving environment and enables these captured features to be replayed in offline training, ensuring perfect consistency between training and serving feature computation in large-scale industrial deployments. 
The configuration-driven approach not only reduces the engineering burden of implementing feature processing pipelines but also ensures reproducibility and consistency across offline training and online serving environments.

The main contributions of this work are as follows:
\begin{enumerate}
    \item We design a unified configuration protocol named \textbf{Prompt Generation} that systematically addresses the heterogeneous feature processing requirements in LLM-based generative retrieval, enabling seamless integration of text, embedding, combo, and sequential features through a declarative interface. The protocol supports flexible feature transformation and compression, allowing developers to control model input complexity and computational cost while maintaining retrieval performance.

    \item We develop a flexible training framework that supports rapid experimentation with different feature combinations and prompt designs through configuration changes alone, significantly reducing the development cycle for building generative retrieval models. This configuration-driven design further enables integration with LLM-driven autoresearch agents for automated feature exploration.

    \item We build an end-to-end training-serving architecture with comprehensive deployment optimizations addressing computational efficiency and system reliability. We implement an event tracking mechanism that logs source features from online serving for offline training replay, ensuring consistency between training and serving environments and eliminating training-serving skew in large-scale industrial deployments.

    \item We conduct extensive experiments on Taobao Search, Taobao Recommendation, and open-source benchmarks, including feature experiments, semantic alignment analysis, and latency analysis. These experiments demonstrate the effectiveness and practical value of our framework in large-scale industrial applications.
\end{enumerate}

The rest of this paper is organized as follows. Section~\ref{sec:protocol} presents the unified configuration protocol and its composable components. Section~\ref{sec:architecture} describes the training and serving architecture, including the event tracking mechanism. Section~\ref{sec:experiments} reports experimental results on Taobao Search, Taobao Recommendation, and open-source benchmarks. Section~\ref{sec:discussion} discusses key findings, integration with autoresearch, and future directions, and Section~\ref{sec:conclusion} concludes.
\section{Protocol Design}
\label{sec:protocol}

\subsection{Design Overview}
\label{sec:protocol_overview}

Our Prompt Generation framework adopts a configuration-driven philosophy that separates feature processing logic from model architecture. At its core, the framework uses a JSON-based configuration schema that declaratively specifies how heterogeneous features should be transformed, composed, and integrated into LLM prompts. This design choice enables rapid experimentation with different feature combinations and prompt structures without modifying the underlying training or serving infrastructure. The configuration schema serves as a universal specification that can be interpreted consistently across both offline training and online inference environments, ensuring feature computation alignment between the two stages.

The design is guided by three key principles. First, flexibility allows developers to easily incorporate new features or modify existing feature processing pipelines through simple configuration changes, eliminating the need for code modifications or redeployment. Second, consistency ensures that features are computed identically during offline training and online serving, preventing training-serving skew that commonly degrades model performance in production systems. This is achieved through shared configuration parsing and feature processing components used in both environments. Third, extensibility enables the framework to accommodate new feature types and processing components as requirements evolve, supporting custom preprocessors, projectors, and mergers through a plugin architecture while maintaining backward compatibility with existing configurations.
A complete end-to-end configuration example integrating all four feature types is provided in Section~\ref{sec:config_example}; readers may find it helpful to review that example before proceeding to the detailed type and component descriptions below.

\begin{figure}[H]
    \centering
    \includegraphics[width=\linewidth]{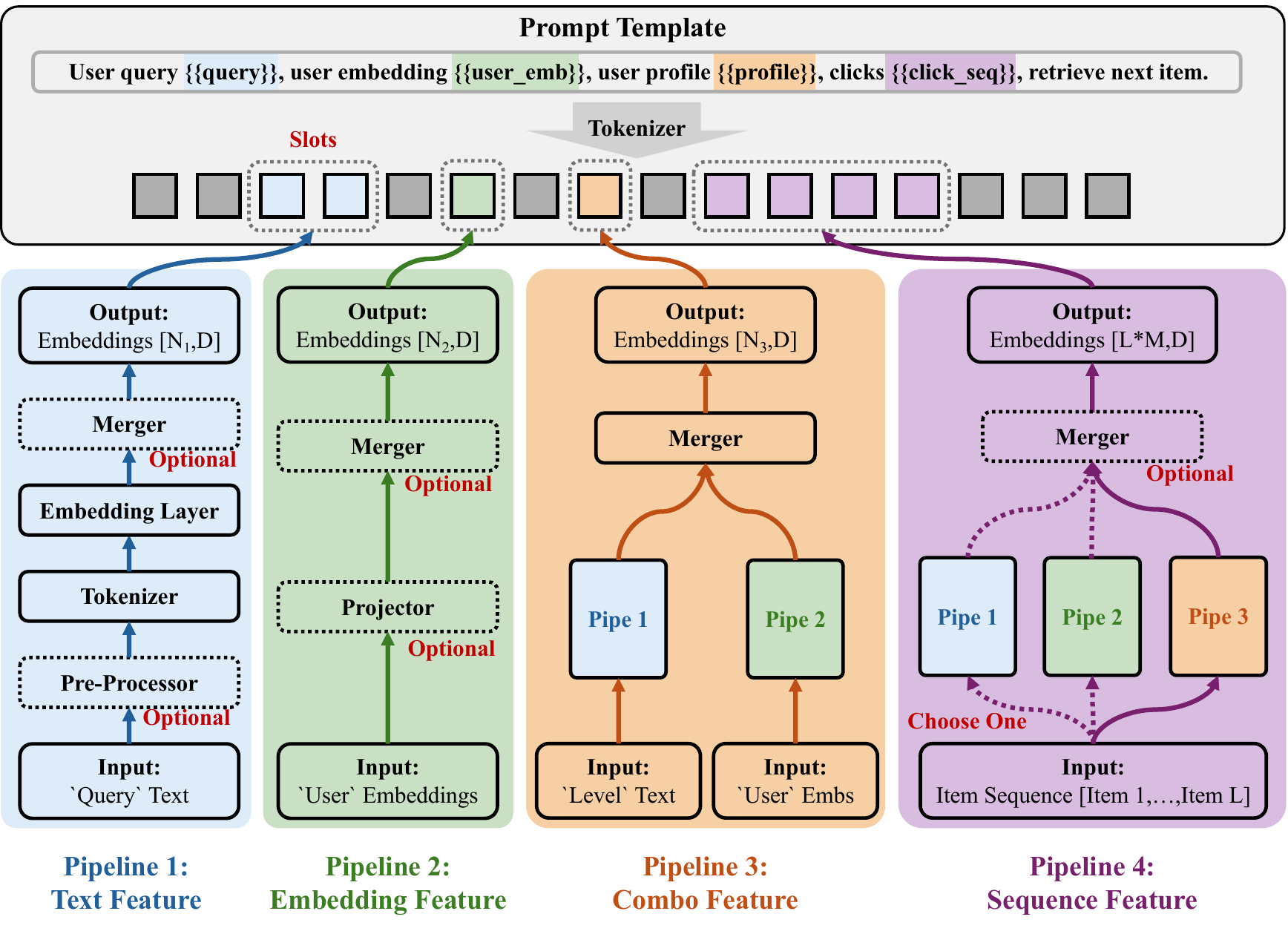}
    \caption{Feature type hierarchy and processing pipeline in Prompt Generation Framework. The framework supports four fundamental feature types: text features (\textcolor{blue}{blue}), embedding features (\textcolor{green}{green}), combo features (\textcolor{orange}{orange}), and sequence features (\textcolor{purple}{purple}). Each pipeline shows the transformation from raw source features to processed embeddings that fill placeholder slots in the prompt template. For sequence features, all items within a sequence must be of the same type (text, embedding, or combo), ensuring consistent processing across the sequence.}
    \label{fig:feature_hierarchy}
\end{figure}

\subsection{Feature Type Taxonomy}
\label{sec:feature_types}

The framework defines four fundamental feature types: text features for raw string inputs, embedding features for pre-computed dense vectors, combo features for composing multiple sub-features, and sequence features for temporal or ordered collections. Figure~\ref{fig:feature_hierarchy} illustrates the type hierarchy and processing pipelines for each category. This taxonomy provides the foundational abstractions for the configuration schema and processing components described in subsequent sections.

\paragraph{Text Features.}

Text features accept raw string inputs that are directly incorporated into the prompt template. After optional preprocessor transformations, the raw strings are tokenized and embedded by the LLM's embedding layer, leveraging the model's learned representations to capture domain-specific semantics through fine-tuning. Typical use cases include:
\begin{itemize}
    \item User queries expressing search intent
    \item Product titles and descriptions
    \item Categorical user attributes (\eg, gender codes preprocessed via mapping to human-readable labels)
\end{itemize}

\paragraph{Embedding Features.}

Embedding features provide pre-computed dense vectors that bypass the LLM's tokenization and embedding layers. Each embedding specifies its shape as a two-dimensional array $[N, D]$, where $N$ is the number of vectors and $D$ the dimensionality. Optional projector components (typically MLPs) align embeddings to the LLM's hidden size or adapt pre-trained representations to the target task. The framework supports both inline embeddings and external table lookups. Common applications include:
\begin{itemize}
    \item User embeddings encoding profiles, preferences, and behavioral patterns
    \item Item embeddings representing product characteristics (visual, textual, or collaborative signals)
\end{itemize}

\paragraph{Combo Features.}

Combo features compose multiple sub-features of any type (text, embedding, or combo) into a single unified representation. Each sub-feature is processed according to its type, then merged using a configured merger strategy. This compositional design supports both simple concatenation and learned fusion through trainable mergers. Common applications include:
\begin{itemize}
    \item Item combo: merging product ID, title, and category into a unified item representation
    \item User combo: combining user attributes, interest tags, and behavioral signals
    \item Context enrichment: augmenting base features with additional contextual information
\end{itemize}

\paragraph{Sequence Features.}

Sequence features represent temporal or ordered collections where each element comprises one or more sub-features of any type (text, embedding, or combo). All elements share the same sub-feature configuration, ensuring uniform representation across the sequence. The framework provides flexible sequence length management through truncation strategies, balancing historical context against computational cost. Common applications include:
\begin{itemize}
    \item Purchase history, Click history and browse sequences
    \item Session-based interaction sequences
\end{itemize}

\subsection{Configuration Schema}
\label{sec:config_schema}

The configuration schema operationalizes the design principles and feature taxonomy through two complementary JSON specifications. The \textbf{feature configuration (prompt\_feature.json)} provides a structured format for declaring features, as well as their types, data sources, and associated processing components including preprocessors, projectors, and mergers. The \textbf{template configuration (prompt\_template.json)} defines the prompt structure with placeholder slots corresponding to configured features, establishing the final text format presented to the LLM.

\subsubsection{Configuration Fields}
\label{sec:config_fields}

\begin{table}[H]
\centering
\caption{Complete configuration field reference for the Prompt Generation framework. Fields are categorized into basic settings, processing components, and control logic.}
\vspace{3mm}
\label{tab:config_fields}
\small
\renewcommand{\arraystretch}{1.4}
\resizebox{\textwidth}{!}{
\begin{tabular}{llp{0.75\linewidth}}
\toprule
\textbf{Field} & \textbf{Required} & \multicolumn{1}{c}{\textbf{Description}} \\
\midrule
\multicolumn{3}{c}{\textbf{Basic Settings}} \\
\midrule
feature\_name & Yes & Placeholder identifier in templates (\texttt{\{\{name\}\}}) and offline training table field name \\
feature\_type & Yes & Feature type: \texttt{text}, \texttt{emb}, \texttt{combo}, or \texttt{sequence} \\
expression & Yes & Data source specification for online serving. \\
dimension & emb & Embedding shape as $[N, D]$ array ($N$ vectors, $D$ dimensions) \\
features & combo/seq & Array of sub-feature definitions \\
sequence\_length & sequence & Maximum sequence elements. Used only in offline training for truncation (no padding) \\
text\_token\_len & text (Opt.) & Fixed token count after tokenization. Required when using certain mergers like concat\_mlp that need fixed-length inputs \\
only\_record & Optional & Feature logged for analysis but not used in model input \\
comment & Optional & Human-readable description for documentation \\
\midrule
\multicolumn{3}{c}{\textbf{Processing Components}} \\
\midrule
pre\_processors & text (Opt.) & Chained preprocessors (mapping, bucketization) before tokenization \\
projector & emb (Opt.) & Dimension alignment module (typically MLP) \\
merger & combo (Req.) & Token aggregation strategy (mean, sum, concat\_mlp). Required for combo features, optional for others \\
\midrule
\multicolumn{3}{c}{\textbf{Control Logic}} \\
\midrule
empty\_processor & Optional & Empty value handling: fill (default) or drop (omit feature). Drop only allowed for top-level text/emb features, not for sub-features in combo/sequence features  \\
prefix\_text & Optional & Text prepended before feature tokens in prompt. Only applies to non-dropped top-level features \\
suffix\_text & Optional & Text appended after feature tokens in prompt. Only applies to non-dropped top-level features \\
\bottomrule
\end{tabular}
}
\end{table}

Table~\ref{tab:config_fields} provides a comprehensive reference of all configuration fields supported in the feature configuration schema. These fields are organized into three categories: basic setting fields that establish feature identity, processing component fields that define transformation and fusion operations, and control logic fields that handle edge cases and formatting. The \textbf{feature\_name}, \textbf{feature\_type}, and \textbf{expression} fields are mandatory for all features, establishing the placeholder identifier and data source. Type-specific fields such as dimension (for embedding features), text\_token\_len (for text features), sequence\_length (for sequence features), and features (for combo and sequence features) are required only when applicable. Processing components including pre\_processors, projector, and merger are optional and can be configured based on specific feature requirements. Control logic fields such as empty\_processor, prefix\_text, and suffix\_text provide fine-grained, optional control over feature behavior and prompt formatting.

The feature\_name serves as the primary identifier, acting as both the placeholder in prompt templates (\texttt{\{\{name\}\}}) and the field name in offline training data tables. To accommodate the disparate requirements of online serving and offline training, we additionally introduce the expression field for online data source specification.
The framework enforces the following constraints on feature\_name to ensure schema integrity:
\begin{itemize}
\item \textbf{Namespace Constraints}: feature\_name must be unique within the same hierarchy level, though it may be reused across different levels.
\item \textbf{Template Scope}: Template placeholders can only reference top-level feature\_name definitions. That is, the sub-features in combo/sequence features cannot be used by template.
\end{itemize}
The mapping between feature\_name and offline data table fields is determined by the feature type:
\begin{itemize}
    \item \textbf{Text and Embedding Features}: These support direct one-to-one mapping, where feature\_name explicitly identifies the corresponding offline table column. 
    \item \textbf{Combo Features}: These act as logical aggregators. The feature\_name is used solely as a template placeholder and does not map to a single offline data table field.
    \item \textbf{Sequence Features}: The outer sequence feature\_name serves as the template placeholder. For sub-features, the framework automatically derives the offline column name by concatenating the outer feature\_name with the sub-feature's feature\_name using an underscore delimiter (e.g., \texttt{click\_seq} + \texttt{\_} + \texttt{title} $\rightarrow$ \texttt{click\_seq\_title}).
\end{itemize}

A complete configuration example including all feature types is provided in Section~\ref{sec:config_example}.

\subsubsection{Processing Components}
\label{sec:schema_processing}

The framework provides three categories of processing components that transform and compose features at different stages of the pipeline. \textbf{Preprocessors} normalize text values before tokenization, \textbf{Projectors} align embedding dimensions, and \textbf{Mergers} aggregate multiple features into unified representations. Each component type can be configured independently per feature, with parameters controlling their behavior as detailed below.

\paragraph{Preprocessors} operate exclusively on text features, transforming raw string values into normalized representations before they are tokenized by the LLM. The framework supports two types of preprocessors, each serving distinct transformation purposes.

\begin{codelisting}[H]
\begin{tcolorbox}
\begin{lstlisting}[style=myminted, language=JSON]
{
    "feature_name": "gender",
    "feature_type": "text",
    "expression": "user:gender",
    "pre_processors": [
        {
            "type": "mapping",
            "params": {
                "dict": {"1": "Male", "2": "Female"},
                "default_value": "Unknown"
            }
        }
    ]
}
\end{lstlisting}
\end{tcolorbox}
\caption{Mapping preprocessor configuration example for transforming gender codes to labels.}
\label{lst:mapping_preprocessor}
\end{codelisting}

\begin{codelisting}[H]
\begin{tcolorbox}
\begin{lstlisting}[style=myminted, language=JSON]
{
    "feature_name": "age",
    "feature_type": "text",
    "expression": "user:age",
    "pre_processors": [
        {
            "type": "bucketize",
            "params": {
                "boundaries": [18, 25, 35, 45, 60]
            }
        }
    ]
}
\end{lstlisting}
\end{tcolorbox}
\caption{Bucketization preprocessor configuration example for discretizing age values into bins.}
\label{lst:bucketize_preprocessor}
\end{codelisting}

\textbf{Mapping Preprocessor} performs value transformations by converting raw values to human-readable labels or canonical forms through a dictionary lookup. It accepts two parameters: \texttt{dict} (required), which defines the mapping from source values to target strings, and \texttt{default\_value} (optional), which specifies the fallback string when a source value is not found in the dictionary. If \texttt{default\_value} is not specified, unmapped values are passed through unchanged. This preprocessor is particularly useful for converting categorical codes into descriptive labels, such as transforming gender codes ``1'' and ``2'' into ``Male'' and ``Female'', or mapping product category IDs to category names. Listing~\ref{lst:mapping_preprocessor} demonstrates a mapping preprocessor configuration for gender code transformation.

\textbf{Bucketization Preprocessor} discretizes continuous numerical values into categorical bins, enabling the model to learn from numerical features through text representations. It requires a single parameter \texttt{boundaries}, which specifies a sorted list of floats defining the bucket boundaries. For boundaries $[a, b, c]$, the preprocessor creates buckets: $(-\infty, a)$, $[a, b)$, $[b, c)$, and $[c, \infty)$, mapping numerical values to bucket indices represented as strings. This transformation is useful for features like age, where grouping into ranges (e.g., ``18-24'', ``25-34'') provides more meaningful categorical representations than raw numerical values. Listing~\ref{lst:bucketize_preprocessor} shows a bucketization preprocessor configuration for age discretization.

Multiple preprocessors can be chained in sequence within the \texttt{pre\_processors} array, with each transformation applied to the output of the previous step, enabling complex pipelines.

\begin{codelisting}[H]
\begin{tcolorbox}
\begin{lstlisting}[style=myminted, language=JSON]
{
    "feature_name": "user_emb",
    "feature_type": "emb",
    "expression": "user:emb",
    "dimension": [1, 256],
    "projector": {
        "type": "mlp",
        "params": {
            "hidden_dimensions": [512, 896],
            "activation": "relu",
            "use_bias": true,
            "dropout": 0.1
        }
    }
}
\end{lstlisting}
\end{tcolorbox}
\caption{MLP projector configuration example for transforming user embeddings.}
\label{lst:mlp_projector}
\end{codelisting}

\paragraph{Projectors} transform embedding features to align their dimensionality with the LLM's hidden size or adapt pre-trained representations to the target task. The framework currently supports MLP-based projection, with extensibility for additional projector types through the plugin architecture.

\textbf{MLP Projector} consists of fully connected layers with activation functions, projecting input embeddings from their original dimension to the model's hidden dimension. It supports several configuration parameters: \texttt{hidden\_dimensions} (optional, default: []), which specifies the dimensions of intermediate hidden layers as an array; \texttt{activation} (optional, default: ``relu''), which defines the activation function; \texttt{use\_bias} (optional, default: true), which controls whether bias terms are included in linear layers; \texttt{dropout} (optional, default: 0.0), which sets the dropout rate for each layer; and \texttt{use\_batch\_norm} (optional, default: false), which enables batch normalization between layers. The projector automatically determines the output dimension based on the model's embedding dimension at initialization. Listing~\ref{lst:mlp_projector} illustrates an MLP projector configuration with three layers (256-dimensional input, two hidden layers of 512 and 896 dimensions, and output to model's embedding dimension).

Projectors are optional for embedding features. When not specified, embeddings must already match the model's hidden dimension and are fed directly to the model without transformation.

\begin{codelisting}[H]
\begin{tcolorbox}
\begin{lstlisting}[style=myminted, language=JSON]
{
    "feature_name": "profile",
    "feature_type": "combo",
    "features": [...],
    "merger": {
        "type": "mean",
        "params": {
            "output_token_len": 1
        }
    }
}
\end{lstlisting}
\end{tcolorbox}
\caption{Mean merger configuration example for averaging feature embeddings.}
\label{lst:mean_merger}
\end{codelisting}

\begin{codelisting}[H]
\begin{tcolorbox}
\begin{lstlisting}[style=myminted, language=JSON]
{
    "feature_name": "user_profile",
    "feature_type": "combo",
    "features": [...],
    "merger": {
        "type": "concat_mlp",
        "params": {
            "output_token_len": 1,
            "hidden_dimensions": [512],
            "activation": "relu",
            "dropout": 0.1
        }
    }
}
\end{lstlisting}
\end{tcolorbox}
\caption{Concat-MLP merger configuration example for learned feature fusion.}
\label{lst:concat_mlp_merger}
\end{codelisting}

\paragraph{Mergers} aggregate multiple features into unified representations, controlling how tokens from different sources are combined. A critical requirement for mergers is that all input features must have identical dimensions before merging. For example, when combining embedding and text features, embeddings must be projected to match the model's hidden dimension through projectors, even when using concat-MLP-based merger strategies. The framework supports multiple merger strategies, each with distinct aggregation mechanisms and configurable output lengths.

\textbf{Mean/Sum Merger} provides two parameter-free aggregation strategies for simple feature fusion. Mean merger computes element-wise averages across feature embeddings, while sum merger performs element-wise addition, both producing single-token representations regardless of input size. These strategies are suitable for simple feature aggregation where equal weighting is appropriate. Both only accept the \texttt{output\_token\_len} parameter, which must be set to 1. Listing~\ref{lst:mean_merger} demonstrates a mean merger configuration.

\textbf{Concat-MLP Merger} concatenates feature embeddings along the token dimension, then applies an MLP to produce a fixed-length output. This learned fusion strategy is more expressive than simple averaging or summation, introducing trainable parameters for adaptive feature combination. It supports several configuration parameters: \texttt{output\_token\_len} (optional, default: 1), which controls the number of output tokens; \texttt{hidden\_dimensions} (optional, default: []), which specifies intermediate layer dimensions; \texttt{activation} (optional, default: ``relu''); \texttt{use\_bias} (optional, default: true); \texttt{dropout} (optional, default: 0.0); and \texttt{use\_batch\_norm} (optional, default: false). Listing~\ref{lst:concat_mlp_merger} demonstrates a concat-MLP merger configuration.

\begin{table}[H]
\centering
\caption{Processing component parameter reference. Parameters are organized by component type, showing which settings are required versus optional along with their purposes.}
\vspace{3mm}
\label{tab:processing_components}
\small
\renewcommand{\arraystretch}{1.5}
\resizebox{\textwidth}{!}{
\begin{tabular}{llp{0.05\linewidth}p{0.55\linewidth}}
\toprule
\textbf{Field} & \textbf{Parameter} & \textbf{Req.} & \multicolumn{1}{c}{\textbf{Description}} \\
\midrule
\multirow{2}{*}{mapping (pre\_process)} & dict & Yes & Dictionary mapping source values to target strings \\
& default\_value & No & Fallback string for unmapped values (default: original string) \\
\midrule
bucketize (pre\_process) & boundaries & Yes & Sorted list of floats defining bucket boundaries \\
\midrule
\multirow{5}{*}{mlp (projector)} & hidden\_dimensions & No & Array of intermediate layer dimensions (default: []) \\
& activation & No & Activation function (default: relu) \\
& use\_bias & No & Include bias terms in layers (default: true) \\
& dropout & No & Dropout rate (default: 0.0) \\
& use\_batch\_norm & No & Enable batch normalization (default: false) \\
\midrule
mean (merger) & output\_token\_len & No & Number of output tokens, must be 1 (default: 1) \\
\midrule
sum (merger) & output\_token\_len & No & Number of output tokens, must be 1 (default: 1) \\
\midrule
\multirow{6}{*}{concat\_mlp (merger)} & output\_token\_len & No & Number of output tokens (default: 1) \\
& hidden\_dimensions & No & Array of intermediate layer dimensions (default: []) \\
& activation & No & Activation function (default: relu) \\
& use\_bias & No & Include bias terms in layers (default: true) \\
& dropout & No & Dropout rate (default: 0.0) \\
& use\_batch\_norm & No & Enable batch normalization (default: false) \\
\bottomrule
\end{tabular}
}
\end{table}

Mergers are required for combo features but optional for other feature types, where they can be used to compress multiple tokens into fewer representations.

The framework provides an extensible plugin architecture that enables developers to implement custom processing components for specialized needs. This extensibility applies to all three component categories through a registration mechanism. For example, developers can implement custom preprocessors for domain-specific text normalization, custom projectors for specialized embedding transformations, or custom mergers such as transformer-based mergers using self-attention mechanisms for context-aware fusion. This design ensures that the framework can accommodate evolving requirements without modifying core infrastructure.

Table~\ref{tab:processing_components} provides a comprehensive reference of all processing component parameters organized by component type.

\subsubsection{Control Logic}
\label{sec:schema_control}

The framework provides control mechanisms that handle missing data and format prompt text. These controls ensure robust prompt generation under diverse data conditions, enabling graceful degradation when features are unavailable while maintaining consistent prompt structure through prefix and suffix text markers.

\begin{codelisting}[H]
\begin{tcolorbox}
\begin{lstlisting}[style=myminted, language=JSON]
{
    "features": [
        {
            "feature_name": "gender",
            "feature_type": "text",
            "expression": "user:gender",
            "empty_processor": {
                "condition": ["__NULL__", "__EMPTY__"],
                "action": "fill",
                "default_value": "Unknown"
            }
        },
        {
            "feature_name": "user_emb",
            "feature_type": "emb",
            "expression": "user:emb",
            "empty_processor": {
                "condition": ["__NULL__", "__EMPTY__"],
                "action": "drop"
            }
        }
    ]
}
\end{lstlisting}
\end{tcolorbox}
\caption{Empty processor configuration examples showing fill action (gender feature with default value) and drop action (user\_emb feature excluded when missing).}
\label{lst:empty_processor}
\end{codelisting}

\paragraph{Empty Value Handling} addresses data absence issues through the \texttt{empty\_processor} configuration, which manages missing or invalid feature values across text, embedding, and sequence feature types. This configuration accepts three parameters: \texttt{condition}, which specifies a list of conditions that trigger empty processing; \texttt{action}, which defines the handling strategy (either ``fill'' or ``drop''); and \texttt{default\_value}, which provides the replacement value when action is ``fill''.

The \texttt{condition} parameter supports two special sentinel values that trigger empty processing based on feature type. ``\texttt{\_\_NULL\_\_}'' matches null or None values across all feature types. ``\texttt{\_\_EMPTY\_\_}'' matches empty strings (zero-length strings) for text features, empty arrays (zero-size arrays) for embedding features, and empty sequences (all sub-features are empty) for sequence features. When a feature value matches any condition in the list, the specified action is triggered. Multiple conditions can be specified simultaneously to handle different types of missing data.

When \texttt{action} is set to ``fill'', missing values are replaced with \texttt{default\_value}, ensuring the feature remains in the prompt. The default value type must match the feature type: strings for text features, and numerical values for embedding features. When \texttt{action} is set to ``drop'', the entire feature is excluded from prompt generation when missing. The drop action is only permitted for top-level features, and sub-features within combo or sequence types must use fill action to maintain dimensional consistency. Listing~\ref{lst:empty_processor} demonstrates both empty processing strategies.

\begin{codelisting}[H]
\begin{tcolorbox}
\begin{lstlisting}[style=myminted, language=JSON]
{
    "feature_name": "query",
    "feature_type": "text",
    "expression": "user:query",
    "prefix_text": "Search query: ",
    "suffix_text": ". "
}
\end{lstlisting}
\end{tcolorbox}
\caption{Prefix and suffix text configuration for structured prompt formatting.}
\label{lst:prefix_suffix}
\end{codelisting}

\paragraph{Prefix and Suffix Text} enable structured prompt formatting by adding descriptive context around features. The \texttt{prefix\_text} and \texttt{suffix\_text} parameters accept arbitrary strings that are prepended and appended to feature content, respectively. These text markers are only applied at the top level. Sub-features within combo or sequence types cannot have their own prefix or suffix text, as this would interfere with hierarchical composition. Both parameters default to empty strings when not specified.

Prefix and suffix text serve multiple purposes in prompt engineering. They provide semantic labels that clarify feature meaning (such as ``User age: '' for age features), maintain grammatical structure through punctuation (such as commas or periods), and create consistent formatting across different feature types. Importantly, these text markers are only included in the final prompt when the feature itself is present. If a feature is dropped due to empty processing, its prefix and suffix are also excluded. Listing~\ref{lst:prefix_suffix} demonstrates structured text formatting for a search query feature, where the prefix provides context and the suffix adds terminal punctuation.

\subsection{Configuration Example Walkthrough}
\label{sec:config_example}

\begin{codelisting}[H]
\begin{tcolorbox}
\begin{lstlisting}[style=myminted, language=JSON]
{
    "prompt": "User query {{query}}, user embedding {{user_emb}}, user profile {{profile}}, clicks {{click_seq}}, retrieve next item.",
    "response": "{{label}}"
}
\end{lstlisting}
\end{tcolorbox}
\caption{Prompt template configuration showing placeholder slots for all features.}
\label{lst:complete_template}
\end{codelisting}

\begin{codelisting}[H]
\begin{tcolorbox}
\begin{lstlisting}[style=myminted, language=JSON]
{
    "features": [
        {
            "feature_name": "query",
            "feature_type": "text",
            "expression": "user:query",
        },
        {
            "feature_name": "user_emb",
            "feature_type": "emb",
            "expression": "user:emb",
            "dimension": [1, 256],
            "projector": {"type": "mlp", "params": {"hidden_dimensions": [512, 896]}}
        },
        {
            "feature_name": "profile",
            "feature_type": "combo",
            "features": [
                {"feature_name": "level", "feature_type": "text", "expression": "user:level", "merger": {"type": "mean"}},
                {"feature_name": "user_emb", "feature_type": "emb", "expression": "user:emb", "dimension": [1, 256], "projector": {"type": "mlp"}}
            ],
            "merger": {"type": "concat_mlp"}
        },
        {
            "feature_name": "click_seq",
            "feature_type": "sequence",
            "sequence_length": 20,
            "features": [
                {"feature_name": "item_id", "feature_type": "text", "expression": "click_seq:item_id"},
                {"feature_name": "item_emb", "feature_type": "emb", "expression": "click_seq:item_emb", "dimension": [1, 128], "projector": {"type": "mlp"}}
            ],
            "prefix_text": "Click history: ",
            "suffix_text": "."
        },
        {
            "feature_name": "label", 
            "feature_type": "text", 
            "expression": "item:item_id"
        }
    ]
}
\end{lstlisting}
\end{tcolorbox}
\caption{Complete feature configuration demonstrating all four feature types (text, embedding, combo, sequence).}
\label{lst:complete_config}
\end{codelisting}

This section presents a concrete e-commerce search configuration example corresponding to Figure~\ref{fig:feature_hierarchy}, which integrates all four feature types. The framework requires two complementary JSON configuration files that work together to define the complete prompt generation pipeline. Listing~\ref{lst:complete_template} shows the prompt\_template.json file, which defines the prompt structure with placeholder slots using double curly brace syntax (\texttt{\{\{feature\_name\}\}}). Listing~\ref{lst:complete_config} shows the corresponding prompt\_feature.json file, which provides detailed configurations for each feature.

The feature name correspondence between the two files operates as follows. Each placeholder in the template (such as \texttt{\{\{query\}\}}) must have a matching top-level feature\_name field in the feature configuration. The framework uses these feature\_name values to identify which feature configurations correspond to which template placeholders, establishing a direct one-to-one mapping. During prompt generation, the framework processes each configured feature according to its type, applies the specified transformations through preprocessors, projectors, and mergers, then inserts the resulting embeddings into the positions marked by the corresponding placeholders in the template. This separation between template structure and feature configuration enables flexible experimentation with different feature combinations and processing strategies without modifying the underlying prompt format.
\section{System Architecture and Implementation}
\label{sec:architecture}

\begin{figure}[H]
    \centering
    \includegraphics[width=\linewidth]{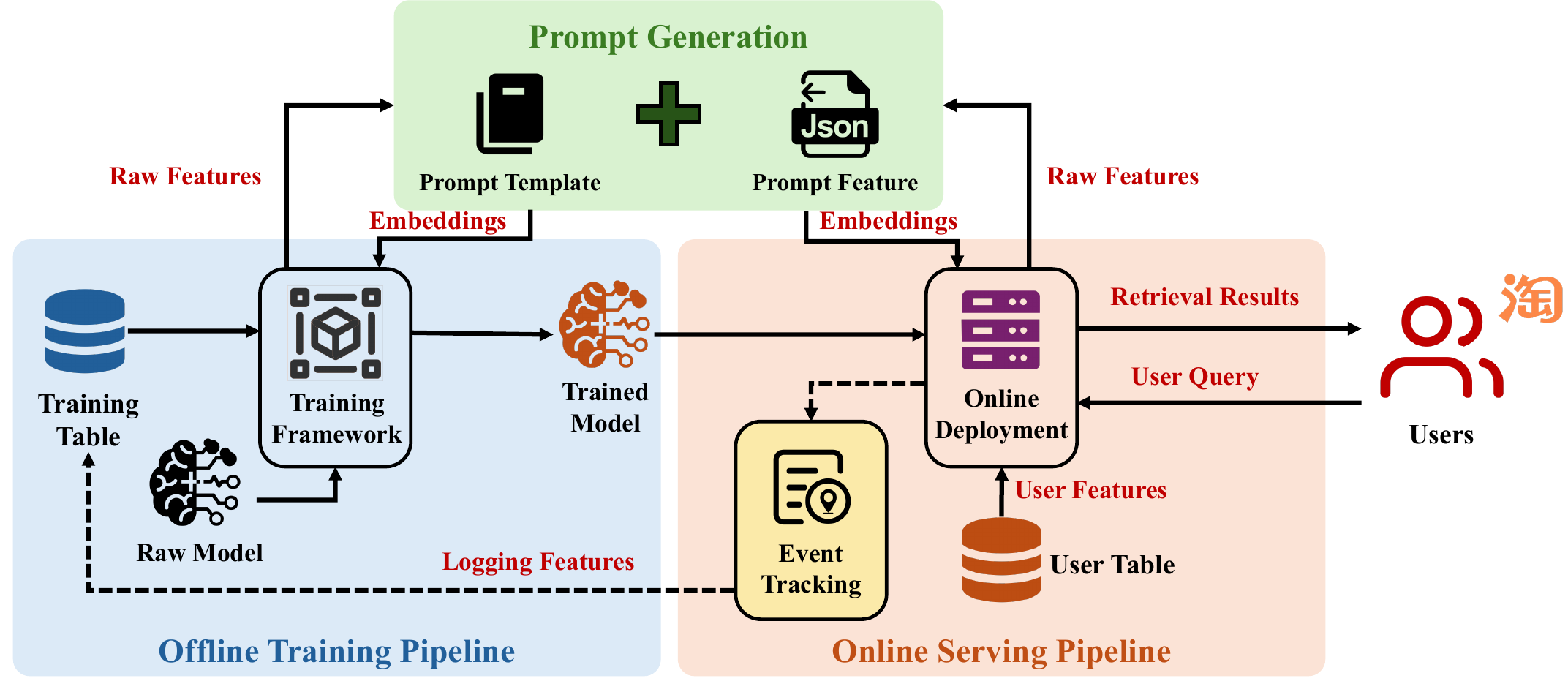}
    \caption{Overall system architecture showing the training-serving closed loop. The system consists of two main pipelines: the offline training pipeline (left) and the online serving pipeline (right). Both pipelines share the same PG configurations to ensure consistency in feature processing between training and inference. The comprehensive event tracking mechanism logs all source features from online serving and feeds them back to training data storage, completing the closed-loop system.}
    \label{fig:system_architecture}
\end{figure}

\subsection{Architecture Overview}
\label{sec:arch_overview}

As illustrated in Figure~\ref{fig:system_architecture}, our system architecture consists of two main pipelines forming a training-serving closed loop. Both pipelines share the same Prompt Generation framework and configuration files to ensure consistency between offline and online environments. In the offline training pipeline, features are extracted from training tables via feature\_name fields, processed through configured preprocessors, projectors, and mergers, and fed into LLMs for training. Conversely, the online serving pipeline retrieves features from real-time data sources (e.g., user profiles) via expression fields, applies the same processing sequence, and generates results using the deployed model.

A key component of this architecture is the comprehensive event tracking mechanism, depicted as the feedback loop in Figure~\ref{fig:system_architecture}. This mechanism logs raw features directly from online serving requests and persists them in training tables, allowing the offline pipeline to replay the exact feature states encountered during inference. This approach ensures that training data remains faithful to the real-world serving distribution, thereby effectively eliminating training-serving skew. Furthermore, this closed-loop design facilitates robust debugging, performance monitoring, and continuous model refinement driven by production traffic.

\subsection{Offline Training Pipeline}
\label{sec:training}

\subsubsection{Data Processing Pipeline}
\label{sec:data_processing}

\begin{figure}[H]
    \centering
    \includegraphics[width=1\linewidth]{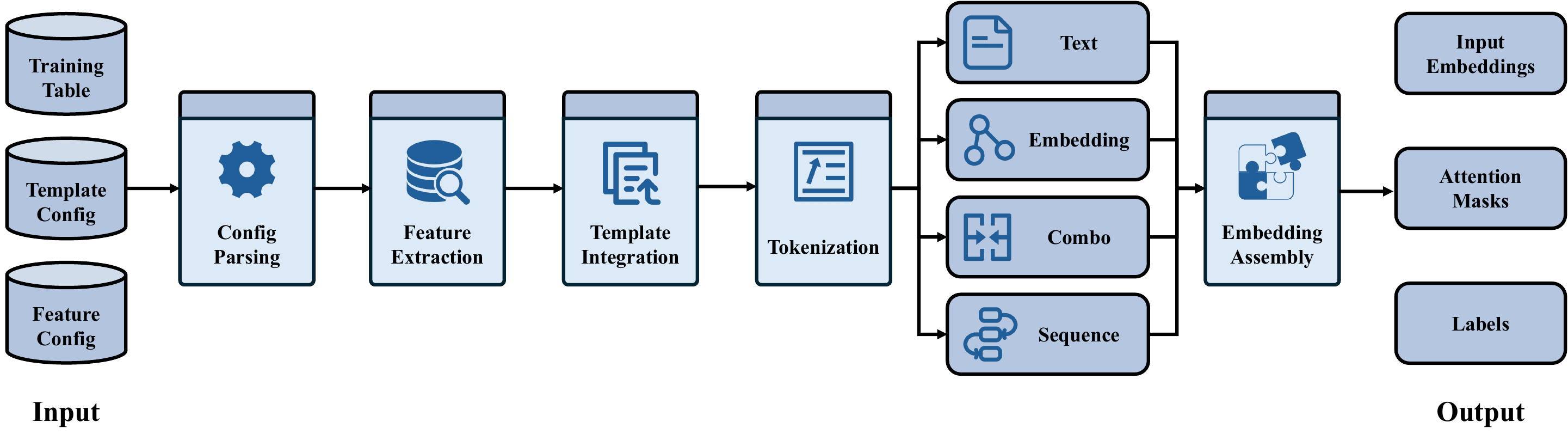}
    \caption{Offline training data processing pipeline. Raw data undergoes multi-stage processing, including config parsing, feature extraction, template integration, tokenization, type-specific transformation, and embedding assembly.}
    \label{fig:training_pipeline}
\end{figure}

The data processing pipeline, illustrated in Figure~\ref{fig:training_pipeline}, transforms raw training data into LLM-compatible embeddings through six coordinated stages:

\begin{enumerate}
    \item \textbf{Config Parsing.} The system loads template and feature configurations from JSON files, constructs a feature processor registry that maps template placeholders to their corresponding processors, and establishes the processing graph.

    \item \textbf{Feature Extraction and Preprocessing.} Raw features are retrieved from training tables based on specified field paths and passed through a dedicated preprocessing pipeline. Initially, empty processors handle missing or null values according to predefined strategies. Subsequently, specific preprocessors normalize and transform raw inputs, including mapping and bucketization.

    \item \textbf{Template Integration with Placeholder Mechanism.} Preprocessed features are integrated into the template, with prefix and suffix text added per feature configuration. Text portions (literal text, prefix/suffix wrappers, text feature values) remain as strings, while non-text feature positions are marked with placeholder identifiers. This produces template slices containing text and placeholders, along with a position-to-feature mapping for subsequent embedding replacement.

    \item \textbf{Tokenization.} Template text slices are tokenized using the base LLM's tokenizer. At placeholder positions, the system infers each feature's output embedding count through its processor, then inserts the corresponding number of placeholder tokens (\texttt{pg\_pad\_id}) to reserve exact space in the token sequence. This ensures sequence length consistency between tokenization and subsequent embedding replacement. Feature values are simultaneously processed by their type-specific processors to prepare for embedding generation.

    \item \textbf{Type-Specific Feature Processing.} The pipeline processes four feature types (text, embedding, combo, sequence) through independent paths, applying type-specific transformations as described in Section~\ref{sec:protocol}. Transformation modules containing learnable parameters (e.g., projection layers, merger networks) are jointly optimized with the base LLM during training.

    \item \textbf{Embedding Assembly.} The final stage scans the tokenized sequence, replacing the placeholder tokens with the corresponding processed feature embeddings. This produces a complete embedding sequence where text tokens and feature embeddings are seamlessly interleaved. For training samples, the pipeline concatenates prompt and response embeddings, generates attention masks, and constructs labels with prompt positions masked.
\end{enumerate}

\subsubsection{Training Framework Integration}
\label{sec:training_framework}

The implementation integrates with HuggingFace Transformers~\citep{wolf-etal-2020-transformers} through a wrapper architecture that adds feature processing capabilities to standard causal language models without modifying their internals.

\paragraph{Model Wrapping Strategy.}
The wrapper maintains two parameter groups with different initialization strategies. Base LLM parameters load from pre-trained checkpoints, preserving language understanding capability. Feature processing parameters (projection layers, merger networks) initialize randomly or from previous prompt generation checkpoints. This design enables a seamless transition to downstream tasks, ensuring that the model benefits from both pre-trained knowledge and task-specific adaptation.

\paragraph{Forward Pass Implementation.}
The system employs a KV-cache inspection mechanism to differentiate between initial prompt processing and subsequent token generation. When the cache is empty, the system identifies the first forward pass and activates the feature-processing layer to replace placeholders with processed feature embeddings and construct the complete input sequence. Conversely, for subsequent generation steps, the system detects a non-empty cache and bypasses the feature-processing layer, forwarding new tokens directly to the LLM. This design ensures that feature processing occurs only once per sample, effectively eliminating overhead during multi-token generation.

\subsection{Online Inference Architecture}
\label{sec:inference}

\subsubsection{Inference Framework Design}
\label{sec:infer_framework}

\begin{figure}[H]
    \centering
    \includegraphics[width=0.8\linewidth]{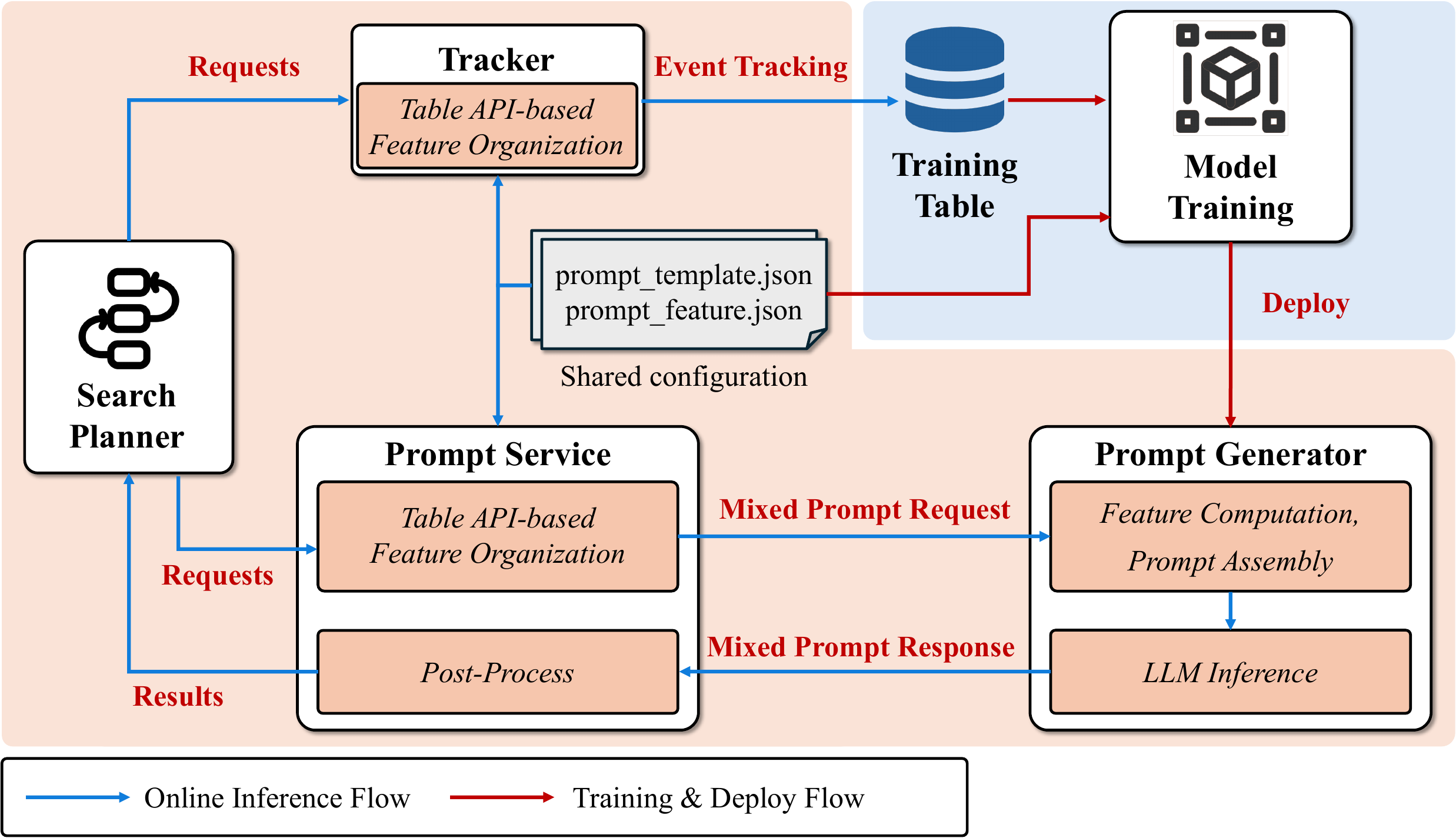}
    \caption{Online inference and event tracking pipeline. Blue arrows trace the flow from \textbf{Search Planner} through \textbf{Tracker}, \textbf{Prompt Service}, and \textbf{Prompt Generator}, with event tracking written back to the Training Table. prompt\_template.json and prompt\_feature.json act as the shared configuration. Red arrows briefly indicate the offline model training and deployment path.}
    \label{fig:online_deployment_tracking}
\end{figure}

\paragraph{Two-layer architecture.}
As illustrated in Figure~\ref{fig:online_deployment_tracking}, the online stack is decomposed into two cooperating components, \textbf{Prompt Service} and \textbf{Prompt Generator}. Prompt Service is the feature organization layer, which performs TableAPI-based feature retrieval, feature preprocessing, mixed input assembly, and post-processes the model output before returning to the caller. Prompt Generator is the prompt computation layer, responsible for tokenization, embedding lookup, module dispatch, final prompt assembly, and then calls the inference backend built on \texttt{RTP-LLM}~\citep{rtp-llm} to perform LLM inference. This separation isolates I/O-bound feature work from compute-bound tensor work, so that each layer can be scaled, batched, and accelerated independently.

\paragraph{Shared request protocol.}
The two layers communicate through a pair of structured messages, \textbf{MixedPromptRequest} and \textbf{MixedPromptResponse}. A MixedPromptRequest carries exactly one prompt template, an ordered list of MixedFeature entries, and a generation configuration. A MixedPromptResponse carries the inference output, which is one of token IDs, decoded text, or an embedding tensor, together with error and trace information. 
\textbf{MixedFeature} is a recursive data type that supports four feature types (TEXT, TOKEN, EMBEDDING, and COMPOUND). Within each MixedFeature, the data payload and the control information are intentionally decoupled so that Prompt Generator can defer operations without blocking the request thread.
The request is interpreted under the same template and feature configurations that drive offline training, thereby ensuring offline-online consistency. 
Online inference additionally introduces a module configuration (\textbf{prompt\_module.json}), automatically exported from the offline trained model and used to carry the learned parameters of mergers, projectors, and other feature-processing modules into the online stack at deployment time.

\paragraph{Inference pipeline.}
Prompt Generator performs the following seven steps:
\begin{enumerate}
    \item \textbf{Request deserialization.} Deserialize all MixedFeature entries to obtain the execution logic.
    \item \textbf{Embedding lookup.} Resolve each MixedFeature into an item embedding according to its type: tokenize-then-lookup for \texttt{TEXT}, direct lookup for \texttt{TOKEN}, the carried embedding for \texttt{EMB}, and recursion on sub-features for \texttt{COMPOUND}.
    \item \textbf{Module dispatch.} Following the dependency order, invoke each MixedFeature's referenced modules, dispatched to predefined operators or to user-defined operators, and producing per-feature embeddings.
    \item \textbf{Template embedding.} Look up the prompt template's token embeddings from the vocabulary table.
    \item \textbf{Placeholder substitution.} Insert the per-feature embeddings produced in step three at the special-token positions of the prompt template, yielding the complete LLM input embedding.
    \item \textbf{LLM inference.} Execute the standard LLM inference procedure.
    \item \textbf{Response assembly.} Return the inference result to Prompt Service via MixedPromptResponse.
\end{enumerate}

To meet the millisecond-level latency budget in production, we further apply the following optimizations along the inference pipeline: (1) memory-pool reuse and torch.compile of merger and projector modules, (2) batched execution of per-feature processors to amortize kernel-launch overhead, (3) parallel multi-stream dispatch and reduced host-side overhead of the beam-search, and (4) beam-search memory-layout optimization. Detailed optimization results and inference efficiency are reported in Section~\ref{sec:exp_latency_infer}.

\subsubsection{Event Tracking}
\label{sec:event_tracking}

In industrial search, advertising, and recommendation systems, user behaviour evolves continuously, so the sequential features retrieved for the same user can differ across requests, which substantially complicates the construction of training tables. A common practice for keeping training and serving features aligned is to filter offline behaviour sequences by the request timestamp, but this approach demands heavy engineering effort and still leaves room for feature misalignment whenever the offline and online preprocessing paths diverge.

To address this, we adopt a full-coverage event tracking pipeline that persists the raw features consumed at serving time back into the training table (Figure~\ref{fig:online_deployment_tracking}). Because online serving and offline training share the same feature and template configurations, the prompts assembled from the replayed features during training are by construction identical to those used online.
In practice, this mechanism achieves over 99\% feature consistency between online serving and offline training.

\section{Experiments}
\label{sec:experiments}

In this section, we evaluate Prompt Generation framework through feature experiments on the Taobao search scenario, the Taobao recommendation scenario, and three open-source e-commerce benchmarks. We then turn to alignment experiments that study how different semantic ID alignment methods interact with PG, and to a latency analysis that covers both offline training and online inference. Finally, we report online A/B results from deploying PG across multiple production search and recommendation scenarios. Our goal here is not to exhaust all feature engineering choices, but to verify that PG can be adapted to \textbf{\emph{multiple base models, multiple application scenarios, and multiple data scales}} through a unified configuration protocol, while preserving competitive retrieval quality.

\subsection{Taobao Search Feature Experiments}
\label{sec:exp_search}

\paragraph{Task Formulation.}
The Taobao search task is to predict the specific product a user will purchase within a given search session. 
We frame this as a generative prediction problem in which the model emits the 3-digit semantic ID (SID) of the target product. 
The input prompt is composed of three structured components, including a textual user profile (\eg, gender), the current search query, and a chronological historical behaviour sequence in which each interaction carries the item SID together with rich contextual signals such as the product title, the shop name, and the interaction timestamp. 

\paragraph{Experimental Setup.}
The dataset is collected from real Taobao search logs, with three consecutive days of transaction samples used for training and a subset of the fourth day used for evaluation. This temporal split simulates a realistic deployment scenario in which the model must generalise to future, unseen traffic. 
We evaluate the model with Hit-Rate at $K$, focusing on HR@50 and HR@5000. HR@50 reflects accuracy on popular head items, whereas HR@5000 measures the model's coverage of long-tail products under a much larger candidate pool. 
We use Qwen2.5-Instruct-0.5B~\citep{qwen2.5} as the backbone fine-tuned to generate the target SID, and we run dynamic beam search~\citep{freitag2017beam} at inference to produce the ranked top-$K$ candidate list. 
To avoid disclosing absolute production metrics, all HR scores in this section are reported as relative gains over the SID-only baseline, which is anchored to 0.00\%.

\paragraph{Offline Results.}
Table~\ref{tab:taobao_search_results} compares eight feature configurations under the same backbone and decoding setup. Adding any single contextual feature (title, shop name, or timestamp) on top of the SID-only baseline brings a small but consistent gain on HR@50, confirming that all three signals carry useful information for ranking the head item. 
However, naively concatenating the textual title roughly $2.7\times$ inflates the prompt token count (from 251 to 667) and lengthens training time correspondingly, while delivering no measurable improvement on HR@5000. 
Replacing raw concatenation with mean-pool mergers keeps the token budget at the same level as the SID-only baseline and preserves most of the per-feature gains. 
Combining the three side features through a combo (mean) operator yields the strongest overall configuration, with the best HR@50 ($+$0.61\%) and the best HR@5000 ($+$0.31\%) gains, while a combo (MLP) variant under the same feature set is slightly weaker. 
The result demonstrates that PG, by exposing both per-feature mergers and multi-feature combinators as configurable components, supports efficient and effective feature exploration without changing model code.

\begin{table}[H]
\centering
\caption{Offline results on the Taobao search scenario. HR@50 and HR@5000 are reported as relative gains (\%) over the SID-only baseline. Blue row denotes the best overall configuration.}
\label{tab:taobao_search_results}
\renewcommand{\arraystretch}{1.3}
\resizebox{\textwidth}{!}{
\begin{tabular}{l c c c c}
\toprule
Configuration & Token Length & Train Time & $\Delta$HR@50 & $\Delta$HR@5000 \\
\midrule
SID only & 251 & 3.88h & 0.00\% & 0.00\% \\
\hdashline
SID + title & 667 & 6.27h & $+$0.26\% & $-$0.04\% \\
SID + shop name & 335 & 5.28h & $+$0.50\% & $+$0.09\% \\
SID + timestamp & 267 & 4.21h & $+$0.23\% & $+$0.23\% \\
SID + mean(title) & 267 & 4.26h & $+$0.15\% & $+$0.25\% \\
SID + mean(shop name) & 267 & 4.19h & $+$0.12\% & $+$0.11\% \\
\rowcolor{bestrow}
SID + combo\_mean(mean(title), timestamp, mean(shop name)) & 267 & 4.65h & \textbf{$+$0.61\%} & \textbf{$+$0.31\%} \\
SID + combo\_mlp(mean(title), timestamp, mean(shop name)) & 267 & 4.68h & $+$0.35\% & $+$0.22\% \\
\bottomrule
\end{tabular}
}
\end{table}

\subsection{Taobao Recommendation Feature Experiments}
\label{sec:exp_rec}

\paragraph{Task Formulation.}
The Taobao recommendation task is to predict the specific product a user will interact with within a single-column session after clicking a trigger product.
Compared to the search task, the recommendation task employs a 2-digit SID for each product.
The input prompt is composed of three structured components, including a textual user profile of user attributes, the SID and contextual features of the current session trigger product (such as category, brand, and seller), and a chronological historical interaction sequence in which each interaction carries the item SID, the item category, the action type (click or purchase), and the elapsed time since the interaction.

\paragraph{Experimental Setup.}
The dataset is collected from real Taobao recommendation logs, with five consecutive days of interaction samples used for training and a subset of the sixth day used for evaluation.
We evaluate the model with Hit-Rate@$K$, focusing on HR@20, HR@200, and HR@2000 to cover head, mid-tail, and long-tail accuracy.
We use Qwen2.5-Instruct-0.5B~\citep{qwen2.5} as the backbone, and we run dynamic beam search~\citep{freitag2017beam} at inference.
To avoid disclosing absolute production metrics, all HR scores in this section are reported as relative gains over the 2-stage mean-pool baseline, which is anchored to 0.00\%.

The Taobao search experiments in Section~\ref{sec:exp_search} have already established that richer feature sets consistently outperform an SID-only configuration. 
We therefore hold the feature set fixed in this section, including the item SID, category, action type, and elapsed time, and study how different ways of composing the same features into the prompt affect downstream performance.
We also extend the PG framework with an \emph{attention} merger that combines sub-features through an attention layer.
This merger is registered through PG's plugin registry without modifying the framework itself, demonstrating that PG natively accommodates user-defined mergers.

\paragraph{Offline Results.}
Table~\ref{tab:taobao_rec_results} compares six feature configurations, with the 2-stage mean-pool as baseline (row 1), which compresses every sub-feature with a per-feature mean and then collapses the four pooled tokens with an outer mean.
The unmerged variant (row 2) delivers the strongest absolute HR gains, with $+$4.32\% on HR@20, $+$3.19\% on HR@200, and $+$1.78\% on HR@2000, but at a $3.4\times$ longer prompt (790 versus 234 tokens).
The single-stage feature-wise pooling variant (row 3) reduces the prompt to 516 tokens and recovers most of the head-item gain, presenting a more balanced compromise.

At the same 234-token budget, replacing the outer mean with a learned combo over fully pooled sub-features (rows 4 and 5) does not consistently improve over the baseline, and both MLP and attention variants regress on HR@20.
Keeping the SID sub-feature unpooled inside the attention combo (row 6) removes this regression and matches the baseline across all three metrics.

No single configuration dominates on both axes, and the optimal choice depends on the available token budget.
PG enables flexible, low-cost experimentation over all such variants, including user-defined mergers, allowing practitioners to select the most cost-effective strategy for their constraints.

Beyond the token budgets compared in Table~\ref{tab:taobao_rec_results}, PG scales to extreme behavior-sequence lengths in this scenario: a configuration whose raw sequence reaches up to 45K tokens, compressed by PG to about 11K tokens, still yields further offline gains, showing that token compression makes otherwise-infeasible ultra-long sequences usable.

\begin{table}[H]
\centering
\caption{Offline results on the Taobao recommendation scenario. HR values are reported as relative gains (\%) over the 2-stage mean-pool baseline (first row). Blue row denotes the best overall configuration.}
\label{tab:taobao_rec_results}
\renewcommand{\arraystretch}{1.3}
\resizebox{\textwidth}{!}{
\begin{tabular}{l c c c c}
\toprule
Configuration & Token Length & $\Delta$HR@20 & $\Delta$HR@200 & $\Delta$HR@2000 \\
\midrule
mean(mean(SID) + mean(category) + mean(time) + mean(action\_type))      & 234 & 0.00\%    & 0.00\%    & 0.00\%    \\
\hdashline
\rowcolor{bestrow}
SID + category + time + action\_type                                    & 790 & \textbf{$+$4.32\%} & \textbf{$+$3.19\%} & \textbf{$+$1.78\%} \\
mean(SID) + mean(category) + mean(time) + mean(action\_type)            & 516 & $+$0.93\% & $+$1.34\% & $+$0.47\% \\
mlp(mean(SID) + mean(category) + mean(time) + mean(action\_type))       & 234 & $-$0.88\% & $+$0.76\% & $-$0.24\% \\
attention(mean(SID) + mean(category) + mean(time) + mean(action\_type)) & 234 & $-$0.72\% & $+$0.44\% & $+$0.08\% \\
attention(SID + mean(category) + mean(time) + mean(action\_type))       & 234 & $+$0.08\% & $+$0.08\% & $-$0.05\% \\
\bottomrule
\end{tabular}
}
\end{table}

\subsection{Open-source Benchmark Feature Experiments}
\label{sec:exp_benchmark}

We evaluate PG on three open-source e-commerce benchmarks with heterogeneous feature structures, semantic ID (SID) formats, and backbone models, namely \textbf{S1-tiny} from FORGE~\citep{fu2025forge}, \textbf{RecIF} from OpenOneRec~\citep{zhou2025openonerec}, and \textbf{Amazon Reviews 2018}~\citep{ni2019justifying} with two domains preprocessed by MiniOneRec~\citep{MiniOneRec}. 
Table~\ref{tab:dataset_stats_open} summarizes the benchmark statistics. Across all benchmarks we adopt dynamic beam search~\citep{freitag2017beam} for decoding, with the beam width progressively expanded to produce a sufficiently large candidate set.

\begin{table}[H]
\centering
\caption{Statistics of the open-source benchmarks.}
\label{tab:dataset_stats_open}
\renewcommand{\arraystretch}{1.3}
\resizebox{\textwidth}{!}{
\begin{tabular}{l l r r l c c}
\toprule
Dataset  & Source & Train & Test & Base Model & SID Tokens & Main Metric \\
\midrule
S1-tiny  & FORGE & 15,697,100 & 9,794 & Qwen2.5-Instruct-0.5B & 3 & HR@K \\
RecIF & OpenOneRec & 113,210 & 27,910 & OneRec-1.7B-pro-pretrain & 5 & Pass@K \\
Amazon (Ind\&Sci)  & MiniOneRec & 36,259 & 4,533 & Qwen2.5-Instruct-0.5B & 3 & HR@K \\
Amazon (Office)  & MiniOneRec & 36,586 & 4,828 & Qwen2.5-Instruct-0.5B & 3 & HR@K \\
\bottomrule
\end{tabular}
}
\end{table}

\subsubsection{S1-tiny}
\label{sec:exp_benchmark_s1}

\paragraph{Experimental Setup.}
S1-tiny is an e-commerce search benchmark from FORGE~\citep{fu2025forge} containing item SID sequences, titles, and multimodal item embeddings. It has roughly 15.7M training examples and 9,794 test examples, with 3 SID tokens per item. We use Qwen2.5-Instruct-0.5B~\citep{qwen2.5} as the backbone with an extended SID vocabulary, and report HR@\{20, 100, 500, 1000\}.

\begin{figure}[H]
    \centering
    \includegraphics[width=\textwidth]{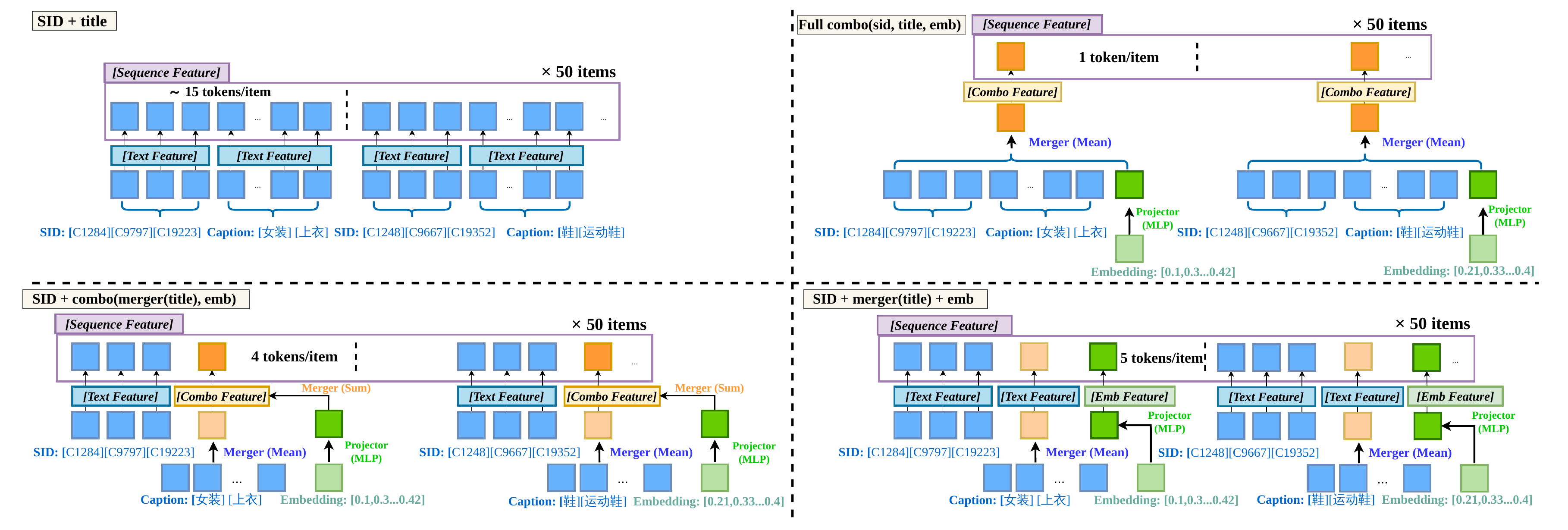}
    \caption{Token-level illustration of representative PG feature configurations on S1-tiny. The figure shows \textit{SID + title}, \textit{Full combo(sid, title, emb)}, \textit{SID + combo(merger(title), emb)}, and \textit{SID + merger(title) + emb}. The comparison highlights how different feature configurations change the number of tokens contributed by each item in the behavior sequence, thereby affecting the trade-off between sequence length and feature expressiveness.}
    \label{fig:s1_tiny_tokenization}
\end{figure}

\paragraph{Results.}
Figure~\ref{fig:s1_tiny_tokenization} illustrates the token-level construction of several representative S1-tiny feature configurations. The figure highlights how PG organizes sequence features under different design choices. The main difference between variants lies in whether textual and embedding signals are directly expanded or compressed through merger modules before entering the final sequence.

Table~\ref{tab:s1_tiny_results} shows that PG can flexibly incorporate text and embedding features through different merger strategies. Adding embedding features yields a small but consistent gain over the SID-only baseline. More importantly, the best result is obtained by first compressing title features and then combining them with embeddings, showing that PG can effectively fuse heterogeneous signals without excessively increasing prompt length. In contrast, fully compressing SID together with text and embedding into a single combo token causes a severe performance drop, indicating that SID should remain an explicit structural signal rather than being overly merged.

\begin{table}[H]
\centering
\caption{Open-source benchmark results on S1-tiny. Blue cells denote the best overall configuration.}
\label{tab:s1_tiny_results}
\resizebox{\textwidth}{!}{
\begin{tabular}{l  c c c c c}
\toprule
Configuration  & Tokens / Item & HR@20 & HR@100 & HR@500 & HR@1000 \\
\midrule
SID only  & 3 & 1.86 & 4.14 & 8.81 & 10.04 \\
\hdashline
SID + title  & $\sim$15 & 1.77 & 4.18 & 8.62 & 9.72 \\
SID + emb  & 4 & 1.83 & 4.22 & 8.91 & 10.16 \\
SID + combo(title, emb) & 4 & 1.81 & 4.30 & 8.84 & 10.08 \\
\rowcolor{bestrow}
SID + combo(merger(title), emb)  & 4 & 1.75 & \textbf{4.41} & \textbf{9.29} & \textbf{10.41} \\
SID + merger(title) & 4 & 1.71 & 4.12 & 8.72 & 9.91 \\
SID + merger(title) + emb  & 5 & 1.58 & 4.40 & 9.00 & 10.35 \\
Full combo(sid, title, emb)  & 1 & 0.19 & 0.73 & 2.37 & 2.97 \\
\bottomrule
\end{tabular}
}
\end{table}

\subsubsection{RecIF}
\label{sec:exp_benchmark_recif}

\paragraph{Experimental Setup.}
RecIF is a cross-domain recommendation benchmark from OpenOneRec~\citep{zhou2025openonerec}. We build an e-commerce-related recommendation subset by selecting samples with valid target annotations and sufficiently rich associated features, yielding about 113K training and 27.9K test examples with 5 SID tokens per item. We use the official OneRec-1.7B-pro-pretrain checkpoint~\citep{zhou2025openonerec}, whose codebook already covers the corresponding SIDs and has acquired some semantic understanding of them through pretraining. We report Pass@\{1, 32, 128\} for this experiment.
Compared with S1-tiny, RecIF has a different SID format, longer sequence types, and a larger backbone model, making it a useful testbed for cross-model and cross-domain adaptability.

\paragraph{Results.}
As shown in Table~\ref{tab:recif_results}, extending the goods SID sequence from length 50 to 100 only brings a marginal improvement, suggesting diminishing returns from simply increasing historical context. This result also highlights one practical advantage of PG, \ie, sequence length can be flexibly configured at the feature level, allowing developers to balance effectiveness and computation cost across different scenarios without modifying the underlying model pipeline. 
Moreover, adding compressed caption text leads to a clear gain, and further introducing video SID information yields the best performance. These results indicate that PG can naturally support cross-domain signals through configuration only, and that compact text compression is a practical way to inject auxiliary semantics without sharply increasing sequence length.

\begin{table}[H]
\centering
\caption{Open-source benchmark results on RecIF. Blue cells denote the best overall configuration.}
\label{tab:recif_results}
\resizebox{0.9\columnwidth}{!}{
\begin{tabular}{l c c c c}
\toprule
Configuration & Seq. Length & Pass@1 & Pass@32 & Pass@128 \\
\midrule
Goods SID only & 50 & 3.13 & 22.90 & 34.70 \\
\hdashline

Goods SID only & 100 & 3.11 & 23.11 & 34.96 \\
Goods SID + merger(caption) & 50 & 3.27 & 24.01 & 35.80 \\
Goods SID + video SID & 50 & 3.24 & 23.03 & 34.90 \\
\rowcolor{bestrow}
Goods SID + merger(caption) + video SID & 50 & \textbf{3.29} & \textbf{24.21} & \textbf{36.34} \\
\bottomrule
\end{tabular}
}
\end{table}

\subsubsection{Amazon Reviews}
\label{sec:exp_benchmark_amazon}

\paragraph{Experimental Setup.}
For Amazon Reviews 2018~\citep{ni2019justifying}, we use two recommendation domains preprocessed by MiniOneRec~\citep{MiniOneRec}, \textit{Industrial\&Scientific} (36,259 train and 4,533 test) and \textit{Office\_Products} (36,586 train and 4,828 test), with 3 SID tokens per item.
The two domains both have relatively small training sets and short user histories. We use Qwen2.5-Instruct-0.5B~\citep{qwen2.5} as the backbone with an extended SID vocabulary, and report HR@\{1, 5, 10, 20, 50\} on both domains.

\paragraph{Results.}
Table~\ref{tab:amazon_results} shows that the optimal feature configuration differs across the two Amazon domains. On Industrial\&Scientific, the gains from adding text features are small overall, and among the tested variants the simple \texttt{+Title} configuration performs the best. On Office, the merger-based \texttt{+merger(Title)+merger(Brand)} configuration is the strongest overall. 
In both domains, several variants underperform the SID-only baseline, indicating that simply adding more text features is not always beneficial. This is likely because the small training set size ($\sim$36K samples) limits the model's ability to reliably leverage additional text signals, making the marginal benefit of extra features unstable. The best feature combination depends on the task, and PG makes it convenient to explore different combinations through configuration changes alone.

\begin{table}[H]
\centering
\caption{Open-source benchmark results on Amazon domains. Left: Industrial\&Scientific. Right: Office\_Products. Blue cells denote the best overall configuration.}
\label{tab:amazon_results}
\renewcommand{\arraystretch}{1.3}
\resizebox{\textwidth}{!}{
\begin{tabular}{l c c c c c | c c c c c}
\toprule
\multirow{2}{*}{Configuration} & \multicolumn{5}{c|}{Industrial\&Scientific} & \multicolumn{5}{c}{Office\_Products} \\
\cmidrule(lr){2-6}\cmidrule(lr){7-11}
& HR@1 & HR@5 & HR@10 & HR@20 & HR@50 & HR@1 & HR@5 & HR@10 & HR@20 & HR@50 \\
\midrule
SID only & 7.21 & 10.08 & 11.54 & 13.66 & 17.23 & 7.52 & 11.27 & \textbf{12.99} & 14.66 & 18.81 \\
\hdashline

+ Title
& \cellcolor{bestleft}\textbf{7.35}
& \cellcolor{bestleft}10.02
& \cellcolor{bestleft}\textbf{11.69}
& \cellcolor{bestleft}\textbf{13.74}
& \cellcolor{bestleft}17.12
& \textbf{8.06} & 11.16 & 12.70 & 14.73 & 18.25 \\
+ Title + Brand & 6.99 & 9.66 & 11.41 & 13.21 & 16.92 & 7.93 & 11.39 & 12.86 & \textbf{15.18} & 18.93 \\
+ merger(Title) + merger(Desc) & 6.95 & 9.55 & 11.27 & 13.66 & \textbf{17.30} & 7.81 & 10.89 & 12.39 & 14.52 & 18.14 \\
+ merger(Title) + merger(Brand)
& 7.08
& \textbf{10.15}
& 11.29
& 13.04
& 16.96
& \cellcolor{bestright}7.60
& \cellcolor{bestright}\textbf{11.47}
& \cellcolor{bestright}12.88
& \cellcolor{bestright}14.95
& \cellcolor{bestright}\textbf{19.37} \\
\bottomrule
\end{tabular}
}
\end{table}

\subsubsection{Summary}
\label{sec:exp_benchmark_summary}

Three main observations emerge from the experiments above.
\begin{enumerate}
    \item Combining features from multiple sources, such as SIDs, text descriptions, and multimodal embeddings, often brings complementary information and improves retrieval quality. 
    \item Directly concatenating raw text features into the prompt is not always the best choice. Mergers offer a useful way to control the prompt token budget while preserving the relevant semantics, and in several settings they also yield more stable results than their uncompressed counterparts.
    \item PG adapts to all three benchmarks despite their different backbones, SID encodings, and task settings, and the above feature combinations are explored purely through PG configuration changes without modifying any model or pipeline code.
\end{enumerate}

\subsection{Alignment Experiments}
\label{sec:exp_alignment}

This section demonstrates that PG's configuration protocol naturally extends to the Pre-SFT alignment stage, where its merger and combo components can be directly reused to design alignment tasks without additional code changes.

\paragraph{Experimental Setup.}
Generative retrieval over semantic IDs requires the model to learn meaningful embeddings for the newly introduced SID tokens before downstream search fine-tuning. 
We instantiate this as a Pre-SFT stage that mixes two complementary families of alignment data: \emph{SID semantic alignment} (\texttt{Item2SID}, \texttt{SID+Merger(Title)2Title}, \texttt{SID2Cate}), which bridges SIDs and item-side information, and \emph{SID semantic retrieval} (\texttt{Query2SID}), which maps user queries to relevant SIDs. 
Among these, \texttt{SID+Merger(Title)2Title} is a reconstruction task that directly optimizes the \texttt{merger} operator, and the sampling ratios of \texttt{Item2SID} and \texttt{Query2SID} are further increased since the downstream objective is SID generation. 
The Pre-SFT stage consumes roughly 300M alignment samples drawn from the four data types listed in Table~\ref{tab:alignment_data}, while the subsequent SFT stage uses training samples collected from 14 days of Taobao search behavioural logs. 
We use \texttt{Qwen2.5-0.5B}~\citep{qwen2.5} as the base model, and the downstream SFT follows the \texttt{SID+Mean(Title)} feature configuration introduced in Section~\ref{sec:exp_search}, so the only difference between the two compared variants is whether the alignment Pre-SFT step is applied. 
We evaluate on two metrics, namely \texttt{search\_pay} (purchase events) and \texttt{search\_ipv} (item page view events), under HR@\{20, 50, 100, 500, 1000, 5000\}.

\begin{table}[H]
\centering
\caption{Statistics of the semantic ID alignment datasets used in the Pre-SFT stage.}
\label{tab:alignment_data}
\renewcommand{\arraystretch}{1.3}
\resizebox{0.8\textwidth}{!}{
\begin{tabular}{l l l l}
\toprule
Data Type & Input & Output & Samples \\
\midrule
\texttt{Item2SID} & Enriched item title & Semantic ID & 100M \\
\texttt{SID+Merger(Title)2Title} & Semantic ID with merged title & Item title & 50M \\
\texttt{SID2Cate} & Semantic ID & Item category & 50M \\
\texttt{Query2SID} & User query & Semantic ID & 100M \\
\bottomrule
\end{tabular}
}
\end{table}

\paragraph{Results.}
Table~\ref{tab:alignment_results} reports the effect of the alignment Pre-SFT stage on the two evaluation metrics. 
To avoid disclosing absolute production metrics, all HR values are reported as relative gains over the \textit{without-alignment} baseline.
Adding the Pre-SFT step yields consistent positive gains on both metrics. Improvements range from $+$1.36\% to $+$3.14\% on \texttt{search\_pay} and from $+$1.63\% to $+$4.56\% on \texttt{search\_ipv}, with the relative gain growing as $K$ increases.
The fact that gains are largest at HR@1000 and HR@5000 indicates that alignment particularly benefits long-tail coverage, broadening the candidate pool the model can recover correctly.
These results show that PG can support semantic-ID alignment as a clean Pre-SFT stage, and that its \texttt{merger} and \texttt{combo} configuration mechanisms remain available inside the Pre-SFT data design without changing the model code.

\begin{table}[H]
\centering
\caption{Effect of the alignment Pre-SFT stage on Taobao search evaluation. Values are HR relative gains (\%) of the alignment setting over the corresponding \textit{w/o alignment} baseline.}
\label{tab:alignment_results}
\renewcommand{\arraystretch}{1.3}
\resizebox{0.9\textwidth}{!}{
\begin{tabular}{l c c c c c c}
\toprule
Slice & $\Delta$HR@20 & $\Delta$HR@50 & $\Delta$HR@100 & $\Delta$HR@500 & $\Delta$HR@1000 & $\Delta$HR@5000 \\
\midrule
\texttt{search\_pay} & $+$1.36\% & $+$1.84\% & $+$2.15\% & $+$2.89\% & $+$2.88\% & $+$3.14\% \\
\texttt{search\_ipv} & $+$1.63\% & $+$2.15\% & $+$2.71\% & $+$3.80\% & $+$4.06\% & $+$4.56\% \\
\bottomrule
\end{tabular}
}
\end{table}

\subsection{Latency Analysis}
\label{sec:exp_latency}

\subsubsection{Training Speed}
\label{sec:exp_latency_train}

\paragraph{Experimental setup.}
We benchmark training speed on a single node with 8 NVIDIA A100-80G GPUs, 128 CPU cores, and 1024~GB of host memory. The base model is Qwen2.5-Instruct-0.5B~\citep{qwen2.5}, and speed is measured in iterations per second (iter/s) together with the wall-clock cost of a single training step.

To cover a representative range of PG configurations, we construct four scenarios drawn from real production usage, varying the number of features, the length of sequences, and the merger and projector mix. 
The configurations are summarised in Table~\ref{tab:training_scenarios}. Scenario~1 is a simple short-sequence configuration, Scenario~2 has medium complexity with several long sequences, Scenario~3 has the most complex feature configuration with the most sequence features and a merger on every sequence sub-feature, and Scenario~4 has the longest LLM input and combines deep nested combo features with a much longer prompt.

\begin{table}[H]
\centering
\caption{Statistics of the four training-speed benchmark scenarios.}
\label{tab:training_scenarios}
\renewcommand{\arraystretch}{1.3}
\resizebox{\textwidth}{!}{
\begin{tabular}{l c c c c c c c}
\toprule
Scenario & \#Feats & \#Text & \#Combo & \#Seq & Seq total len & \#Merger / projector & Avg LLM input \\
\midrule
S1 (simple, short sequence) & 8  & 6  & 0 & 2  & 100  & 5   & 205 \\
S2 (medium, multi long seq.)    & 10 & 5  & 0 & 5  & 1200 & 10  & 794 \\
S3 (complex, sequence-heavy)    & 23 & 9  & 0 & 14 & 1304 & 84  & 1093 \\
S4 (deep nested combo, long input)  & 32 & 20 & 5 & 7  & 105  & 116 & 3260 \\
\bottomrule
\end{tabular}
}
\end{table}

\paragraph{Performance challenges.}
We use Scenario~3, the most sequence-heavy configuration, as the optimization target. 
The baseline implementation suffers from two main bottlenecks.
\begin{itemize}
    \item \textbf{List/dict feature storage.} Intermediate feature values are stored in Python list and dict containers, which triggers deep-traversal hooks inside accelerate, Transformers, and PyTorch DDP and inflates per-step overhead.
    \item \textbf{Per-feature fragmented execution.} Sequence and combo features are assembled feature by feature, and mergers and projectors are invoked per feature. Each individual operation is too small to amortise its kernel-launch overhead, so the GPU spends most of its time launching kernels rather than computing.
\end{itemize}

\begin{figure}[H]
    \centering
    \includegraphics[width=0.8\linewidth]{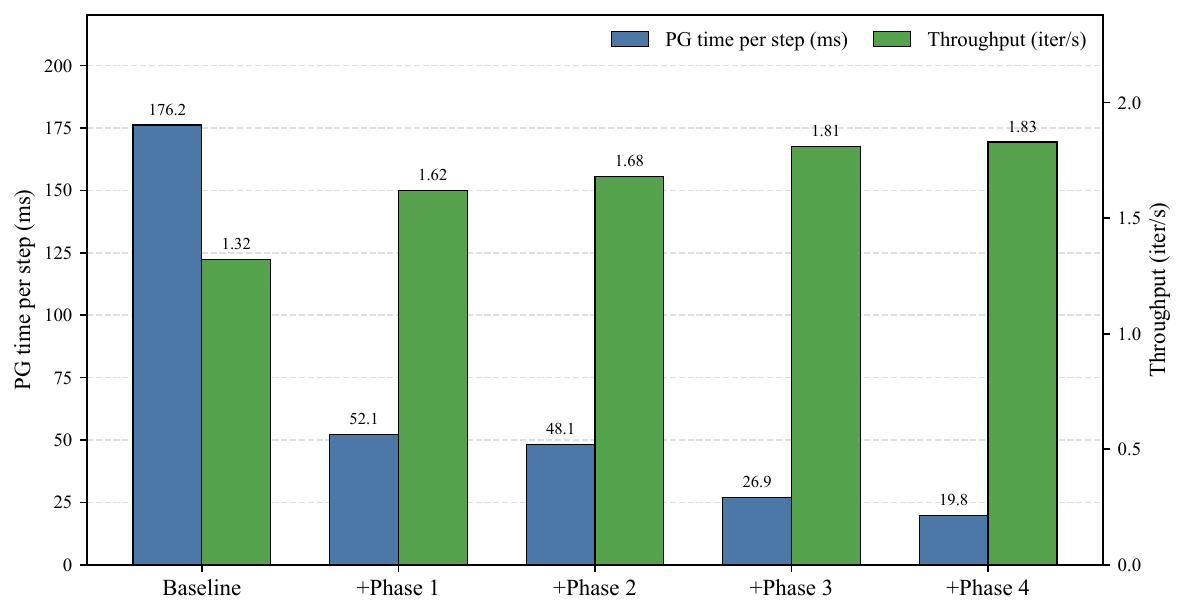}
    \caption{Per-step PG time and throughput on Scenario~3 across optimization phases. Left axis: PG time (ms, lower is better). Right axis: throughput (iter/s, higher is better).}
    \label{fig:training_speed_opt}
\end{figure}

\paragraph{Performance optimization.}
We attack the bottlenecks above in four progressive phases, each addressing a separate group of issues.
\begin{enumerate}
    \item \textbf{Phase 1: traversal-safe feature container.} We replace Python list and dict feature storage with two custom container classes that are not recognised by the deep-traversal hooks in accelerate, Transformers, and PyTorch DDP, eliminating the per-step traversal overhead.
    \item \textbf{Phase 2: batched sequence and combo assembly.} We flatten an entire batch of sequence and combo features into a single embedding tensor and use one index tensor to perform the assembly in one shot, replacing per-feature assembly with a single grouped operation.
    \item \textbf{Phase 3: layered batched merger and projector execution.} We group all same-type merger and projector invocations at the same nesting level into a single batched call, amortising kernel-launch overhead across features that share an operator.
    \item \textbf{Phase 4: C++ index-tensor construction.} We reimplement the index-tensor construction in C++ to remove the residual Python-side overhead in the assembly path.
\end{enumerate}

Figure~\ref{fig:training_speed_opt} reports the per-step PG time and the end-to-end training throughput across the four phases. PG time drops from $176.2$~ms at the baseline to $19.8$~ms after Phase~4, an $88.8\%$ reduction, while throughput rises from $1.32$~iter/s to $1.83$~iter/s ($+38.6\%$) and GPU utilization moves from $63\%$ to $86\%$. After optimization, the PG-related stages collectively account for only $3.6\%$ of the total step time, with the remaining cost shifted to the LLM forward and backward passes.

\paragraph{Per-scenario overhead.}
For each scenario, we compare PG+LLM training against a pure-LLM baseline. Both variants share the same training arguments, LLM input length, and runtime environment. Table~\ref{tab:training_scenario_overhead} reports the per-step time of both variants, the absolute PG time, and its share of the total step time. PG adds $5.6$~ms to $74.1$~ms per step, corresponding to $3.2\%$ to $15.1\%$ of total step time, which is acceptable in production training.

Two patterns emerge from the comparison. 
First, the batched-execution optimization eliminates the cost of mergers and projectors. Scenario~3 has $8.4\times$ more mergers than Scenario~2 (84 versus 10), yet its PG time grows by only $9\%$ (from $67.9$ to $74.1$~ms). 
Second, as the LLM input grows, the PG share shrinks. Scenario~4 has the longest LLM input ($3260$ tokens), and its PG share is only $4.5\%$ despite a non-trivial absolute cost.

\begin{table}[H]
\centering
\caption{Per-step PG overhead across the four training scenarios. Pure LLM denotes a baseline that uses identical training arguments and LLM input length but skips all PG-side feature processing.}
\label{tab:training_scenario_overhead}
\renewcommand{\arraystretch}{1.3}
\resizebox{\textwidth}{!}{
\begin{tabular}{l c c c c c c}
\toprule
Scenario & Batch size & Avg LLM input & Pure LLM step (ms) & PG+LLM step (ms) & PG time (ms) & PG share \\
\midrule
S1 & 10 & 205  & 166.7 & 172.3 & 5.6  & 3.2\% \\
S2 & 10 & 794  & 379.8 & 447.7 & 67.9 & 15.1\% \\
S3 & 10 & 1093 & 470.1 & 544.2 & 74.1 & 13.6\% \\
S4 & 5  & 3260 & 614.2 & 643.6 & 29.4 & 4.5\% \\
\bottomrule
\end{tabular}
}
\end{table}

\subsubsection{Inference Speed}
\label{sec:exp_latency_infer}

\paragraph{Experimental setup.}
We benchmark inference performance under the Taobao Newdetail (ND) scenario, one of the core recommendation scenarios in the Home Guess product. A single ND request consists of a generative recommendation query with 12 placeholder features arranged in a nested compound structure, followed by a variable-width beam search (128 beams in the first step, expanding to 1024 in the second) with {max\_new\_tokens}=2.
All experiments use Qwen2.5-0.5B~\citep{qwen2.5} as the base model, deployed on a single NVIDIA H20 GPU with BF16 precision.
Latency and throughput are measured under steady-state load and reported as P50 and P99 end-to-end response time (RT) together with the sustainable queries-per-second (QPS).

\paragraph{Performance challenges.}
Before any optimization is applied, the inference pipeline exhibits several performance bottlenecks across both the Python feature-processing layer and the C++ inference backend. The main bottlenecks are summarized as follows.
\begin{itemize}
    \item \textbf{GPU kernel fragmentation in Prompt Generator.} The request graph contains dozens of small per-feature processors whose individual workload is far smaller than the kernel-launch overhead, so the GPU spends most of its time launching kernels rather than computing.
    \item \textbf{Frequent CUDA memory allocation.} Intermediate tensors are allocated and freed on every request, and cudaMalloc costs accumulate quickly across the many small processors.
    \item \textbf{Python dispatch overhead.} Under eager mode, the Python-to-C++-to-CUDA dispatch chain dominates the runtime of small modules whose actual compute is only a few microseconds.
    \item \textbf{Beam-buffer over-allocation in RTP-LLM.} The variable-width beam search inflates the active beam count from 128 to 1024 between two decoding steps, which forces worst-case allocation of beam-related buffers far beyond the actual generated length.
    \item \textbf{Serial per-stream dispatch in RTP-LLM.} At 1024 beams, dispatching streams one by one turns the backend into a CPU-bound bottleneck under high concurrency.
\end{itemize}

\paragraph{Performance optimization.}
We attack the bottlenecks above in four progressive phases, each addressing a separate group of issues.

\begin{enumerate}
    \item \textbf{Phase 1: tensor reuse and Python dispatch elimination.} We introduce a CUDA tensor pool that recycles intermediate buffers across requests, apply \texttt{torch.compile} to the merger and projector modules to remove Python-side dispatch overhead, and replace tensor concatenation in the concat\_mlp path with slice-assignment to avoid extra allocations.
    \item \textbf{Phase 2: batched feature processing.} We restructure the per-feature execution into a four-stage pipeline (\emph{build trees}, \emph{collect leaves}, \emph{batch process}, \emph{process trees}) so that the dozens of small processor invocations are fused into a small number of grouped GPU kernels, reducing kernel-launch overhead.
    \item \textbf{Phase 3: parallel stream dispatch and lighter beam-search host runtime.} Inside RTP-LLM we parallelize per-stream dispatch with a thread pool, cache sampler host buffers, and templatize the KV-cache management path, which removes the CPU bottleneck observed at 1024 beams.
    \item \textbf{Phase 4: beam-search memory layout.} We replace worst-case pre-allocation of beam-search token buffers with on-demand allocation sized to the actual generated length, and skip unnecessary beam-logits copies, further reducing both memory traffic and dispatch cost.
\end{enumerate}

\begin{figure}[H]
    \centering
    \includegraphics[width=0.8\linewidth]{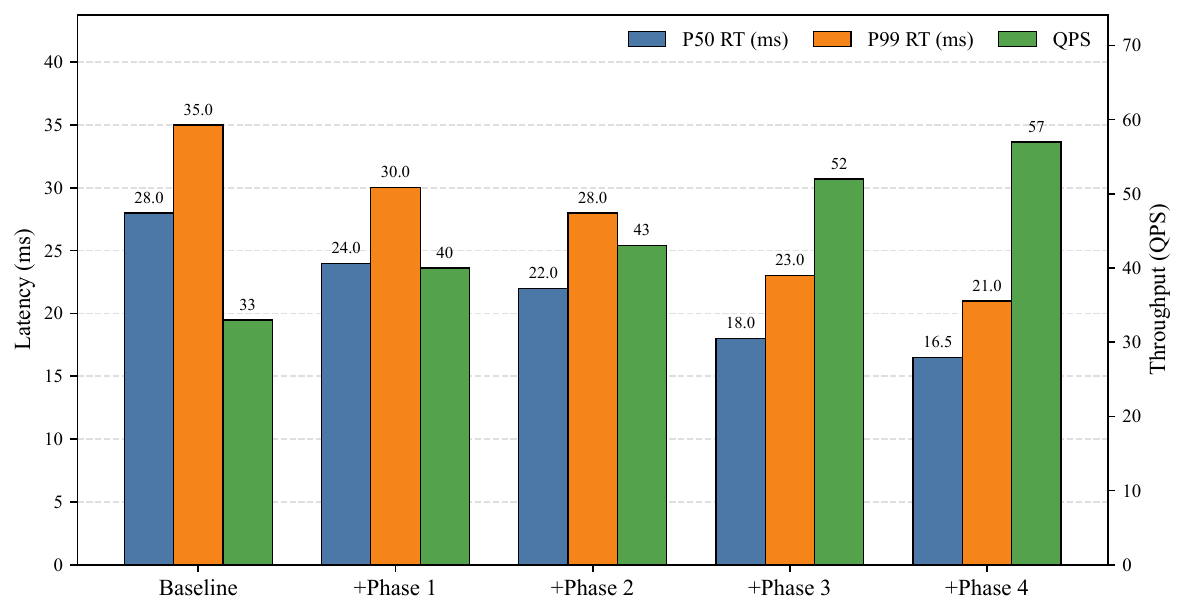}
    \caption{End-to-end inference performance under the ND workload across optimization phases. Left axis: P50/P99 latency (ms, lower is better). Right axis: sustainable QPS (higher is better).}
    \label{fig:inference_speed_opt}
\end{figure}

Figure~\ref{fig:inference_speed_opt} reports P50 RT, P99 RT, and sustainable QPS after each optimization phase. The four phases compound to a substantial improvement, with end-to-end P50 RT dropping from 28.0~ms to 16.5~ms ($\downarrow$41\%), P99 RT from 35.0~ms to 21.0~ms ($\downarrow$40\%), and sustainable QPS rising from 33 to 57 ($\uparrow$73\%).
Overall, Phases 1 and 2 reshape the Python feature-processing path to be memory- and dispatch-friendly, while Phases 3 and 4 reshape the C++ backend around the dynamically growing beam-search workload. The combined optimizations significantly reduce service latency and raise the supportable QPS, allowing the generative recommendation system to serve production traffic at lower hardware cost and leaving sufficient headroom for larger beam widths and more complex feature configurations.

\subsection{Online Deployment}
\label{sec:exp_deployment}

Beyond offline evaluation and micro-benchmarks, PG has been applied in production across multiple Taobao search and recommendation scenarios. In each scenario, adopting PG requires only authoring the two declarative configuration files, \texttt{prompt\_template.json} and \texttt{prompt\_feature.json}, and registering any scenario-specific mergers through the plugin registry; the same configuration then drives both offline training and online serving, so no scenario-specific code is added to the serving path, and the inference-side optimizations of Section~\ref{sec:exp_latency_infer} apply directly. An event-tracking mechanism logs the source features from online serving and replays them during offline training, keeping features consistent across the two stages. Each scenario is deployed on the hardware of its own production cluster, so the serving hardware differs across deployments. We now detail how PG was deployed in several representative scenarios and the online gains it delivered in each. 

\paragraph{Taobao Search.}
We deploy PG on Taobao Search using the same Qwen2.5-0.5B backbone as in Section~\ref{sec:exp_search}, where PG assembles a rich feature configuration and compresses it from over $1{,}400$ tokens to an average of $610$ tokens before it enters the model. PG is introduced as a new retrieval channel that unifies recall and coarse ranking, with the model's generated candidates passed directly to the fine-ranking stage. In an online A/B test on 1\% of search traffic over 14 days, the PG channel increases overall transaction count by $+0.47\%$ and GMV by $+0.51\%$ over the same production system without the PG channel. At the retrieval-channel level, the PG channel raises its share of transactions by $+12$pp and its exclusively-captured transactions (those not covered by any other channel) by $+0.8$pp, showing that it contributes substantial and largely non-overlapping coverage. The simultaneous lifts on count and GMV indicate that these gains reflect both higher user conversion and greater commercial impact on a large-scale production search system. The service runs on NVIDIA H20 GPUs with a $60$\,ms end-to-end response time.

\paragraph{Taobao Recommendation.}
We deploy PG on the Newdetail (ND) scenario of Home Guess, where the baseline is a generative-retrieval model that runs without PG on an input of roughly 800 tokens. PG lets the model consume a much richer raw feature set of over 4{,}000 tokens, while its built-in token compression folds this back to an average of 800 tokens, keeping the serving-time input at essentially the same length as the no-PG baseline. In an online A/B test on 2\% of ND traffic over 12 days, the PG configuration improves ND IPV (item page views) by $+0.66\%$ and PVR by $+7.93\%$ over the no-PG baseline. This deployment runs in the production serving environment on AMD Instinct MI308X GPUs, distinct from the controlled H20 benchmark of Section~\ref{sec:exp_latency_infer}, where PG raises the end-to-end response time only modestly, from 52\,ms to 63\,ms.

\paragraph{Shop Search.}
We further apply PG to Shop Search, which is served by an in-house base model, TBStars-3B. PG assembles a rich feature configuration and compresses the raw input from over $1{,}000$ tokens to an average of $430$ tokens before it enters the model. In an online A/B test on 10\% of Shop Search traffic over more than two weeks, the PG configuration increases transaction count by $+4.01\%$ over the same generative-retrieval model without PG. The service runs on NVIDIA L20 GPUs with an $80$\,ms end-to-end response time.

\paragraph{Other Scenarios.}
Beyond the deployments and gains reported above, PG has also been adopted as a general framework for iterative GR development in a wide range of other applications. These include the the homepage ``Guess You Like'' product recommendation model, content search models, product-detail-page recommendation models, and new-item recommendation models. The broad adoption of PG across these heterogeneous scenarios further demonstrates its generality, flexibility, and effectiveness as a unified framework for generative retrieval.

\section{Discussion}
\label{sec:discussion}

\subsection{Key Findings}
\label{sec:findings}

We obtain three key findings from the above experiments.
\begin{enumerate}
    \item \textbf{There is no universal feature recipe for generative retrieval.} Across Taobao Search (Section~\ref{sec:exp_search}), Taobao Recommendation (Section~\ref{sec:exp_rec}), and the three open-source benchmarks (Section~\ref{sec:exp_benchmark}), combining the item SID with auxiliary signals consistently helps, but the best configuration differs by scenario. PG is designed to make this kind of scenario-dependent exploration cheap.

    \item \textbf{Heterogeneous signals deserve heterogeneous handling.} Different feature types respond differently to the same processing component, so a single one-size-fits-all merger rarely works well. In our experiments, text tolerates heavy mean-pool compression with little loss, whereas the SID collapses retrieval quality when merged together with text and embeddings into one combo token. The general principle is to pick processing components per feature type, which PG's per-feature configuration makes straightforward.

    \item \textbf{The parameter-free mean-pool merger is a strong default.} At a fixed prompt budget, mean-pool consistently matches or outperforms its learned counterparts (MLP, attention) on Taobao Search, Taobao Recommendation, and S1-tiny, since adding learned parameters on top of heavily compressed inputs brings limited headroom. Learned mergers should be reserved for cases whose content-dependent fusion is clearly motivated.
\end{enumerate}

\subsection{Design Insights}
\label{sec:insights}

The central design decision in PG is to separate feature-processing logic from model architecture and to expose the entire feature surface through two declarative JSON files, \texttt{prompt\_template.json} and \texttt{prompt\_feature.json}, which serve as the single source for both offline training and online inference. 
In industrial generative retrieval, the feature set and feature composition change far more often than the model backbone. 
Moving these decisions into configuration turns every feature-engineering iteration into a lightweight edit. 
It also keeps offline training and online inference aligned at the schema level, since both consume the same configuration. 
Together with the event-tracking mechanism in Section~\ref{sec:event_tracking}, this lets the offline pipeline replay the exact features seen online. 

A second design decision keeps the feature taxonomy narrow at four types: text, embedding, combo, and sequence, and factors all transformations into three orthogonal component axes: preprocessors, projectors, and mergers.
These primitives cover the heterogeneous feature mixes seen in most industrial production.
We keep the abstraction narrow because every new type or component has to be supported simultaneously in offline training and online inference, and type explosion directly inflates engineering and validation cost.
Notably, among the three types of components, the merger plays a dual role. Beyond aggregating sub-features, it also controls how many tokens each composed feature contributes to the prompt, so that rich textual features can be incorporated without inflating sequence length.
Because merger strategies live in configuration rather than the model, the trade-off between information density and sequence length can be tuned per feature without modifying code.

\subsection{Integration with Autoresearch}
\label{sec:autoresearch}

\emph{Autoresearch}, where an LLM-driven agent autonomously runs the iterative experimentation loop, has recently emerged as a promising direction for amortising the experimentation cost of large-scale model tuning~\citep{liu2026autoresearchclawselfreinforcingautonomousresearch}. The agent proposes hypotheses, modifies the code, executes experiments, evaluates the results, and iterates toward the user-specified objective with minimal human intervention.

PG framework fits this paradigm particularly well on the feature engineering axis. Adding, removing, or reorganising features in a conventional training pipeline typically requires invasive edits across data loading, tokenization, and input assembly, all of which are difficult for an agent to modify safely and consistently. With PG, the same operations reduce to local edits in prompt\_template.json and prompt\_feature.json, after which the training stack picks up the new configuration without any code change. We validated this combination on the Taobao generative retrieval scenario by integrating PG with an internal autoresearch framework. Starting from the production baseline, the agent explored feature selections, combinations, and prompt template variants through only configuration edits, improving hitrate@20 by $+$3.05\% over 20 iterations.

\subsection{Future Work}
\label{sec:future}

We plan to extend PG to support image as a new feature type.
Recent industrial search, advertising, and recommendation systems increasingly incorporate raw visual content to improve retrieval performance, as pre-encoded image embeddings often fail to preserve fine-grained product information.
A natural starting point is to treat a vision encoder's patch embeddings as a sequence of embedding features, reusing the existing projector and merger infrastructure to compress visual tokens.

We also plan to enrich the merger library with operators that model cross-feature interaction.
The current mean and sum mergers ignore interaction entirely, while concat-MLP captures it only through a fixed feed-forward layer.
Attention-based or low-rank cross-feature mergers would give the framework a more expressive way to compose sub-features without expanding the feature taxonomy.

In addition, we plan to deepen the integration with autoresearch agents, enabling automated exploration over not only feature configurations but also merger strategies and sequence length trade-offs.

\section{Conclusion}
\label{sec:conclusion}

We presented Prompt Generation (PG), a configuration-driven framework for LLM-based generative retrieval that turns feature engineering into a declarative process. 
The PG framework consists of two parts. 
A unified configuration protocol (Section~\ref{sec:protocol}) covers heterogeneous feature processing in a declarative form. 
A training and serving architecture (Section~\ref{sec:architecture}) drives both pipelines from the same configuration, with an event-tracking mechanism used to eliminate training-serving skew. 
By making feature processing declarative, PG simplifies the feature experimentation workflow and lets developers conveniently explore the impact of different feature combinations on the model. 
In the offline feature studies (Section~\ref{sec:experiments}), configuration changes alone yield a series of feature combinations that surpass the baseline. 
Online A/B testing on Taobao Search further confirms statistically significant uplifts on both transaction count and GMV. 
In addition, a series of training and inference optimizations (Section~\ref{sec:exp_latency}) reduce the overhead introduced by PG to a negligible level, making PG a practical framework for industrial-scale generative retrieval.
Despite these results, our evaluation has focused exclusively on e-commerce scenarios; the generalizability of PG to other domains such as content recommendation and sponsored advertising remains to be validated. In addition, the current set of merger operators is limited, and more expressive fusion strategies may be needed for scenarios that require complex cross-feature interactions.

\phantomsection
\addcontentsline{toc}{section}{References}
\bibliography{main}

\newpage
\phantomsection
\addcontentsline{toc}{section}{Contributions}
\section*{Contributions}

\begin{multicols}{2}
\noindent
\textbf{Core Contributors} \\
Dan Ou \\
Gui Ling \\
Hao Wan \\
Hongbin Zhou \\
Jialiang Cheng \\
Jiangnan Pang \\
Silu Zhou \\
Wei Shi \\
Weichen Ye \\
Wenming Zhang \\
Yang Wang \\
Yu Li \\
Yuliang Yan \\
$\text{Zhan Fa}^{*}$ \\
Zhihong Chen \\
Zongyuan Wu \\

\columnbreak

\noindent
\textbf{Contributors} \\
Bo Zheng \\
Changfa Wu \\
Dunxian Huang \\
Haihong Tang \\
Jinlong Guo \\
Kaixuan Zhang \\
Kun Ma \\
Lin Qu \\
Longbo Zhong \\
Tao Lan \\
Tong Xiong \\
Zhibo Wu \\
\end{multicols}

\noindent
$*$ Nanjing University.

\noindent
The listing of authors is in alphabetical order based on their first names.

\vspace{0.5em}
\noindent
We thank all Contributors above for providing experimental support and for many valuable discussions.

\end{document}